\documentclass[a4paper]{article}

\usepackage[english]{babel}
\usepackage{amsmath,amsthm,amssymb}
\usepackage{authblk}
\usepackage{fullpage}

\addto\captionsenglish{}
\usepackage[labelfont=bf]{caption}

\usepackage{cite}
\usepackage{subfig}
\usepackage{hyperref}
\usepackage{tikz}
\usepgflibrary{arrows}
\usepackage{paralist}

\usetikzlibrary{decorations.pathmorphing}
\usetikzlibrary{decorations.text}
\usetikzlibrary{decorations.pathreplacing} 
\usepgflibrary{decorations.footprints}
\usetikzlibrary{shapes}

\usepackage{comment}
\usepackage[ruled, linesnumbered]{algorithm2e}
\usepackage{setspace}
\usepackage[section]{placeins}

\usepackage{hyperref}
\usepackage{enumerate}
\usepackage{mathtools}

\newtheorem{example}{Example}
\newtheorem{theorem}{Theorem}

\newtheorem{lemma}{Lemma}

\newtheorem{claim}{Claim}
\theoremstyle{remark}
\newtheorem{remark}{Remark}
\usepackage{graphicx}

\newcommand{\Nats}{\mathbb{N}}
\newcommand{\Natsplus}{\mathbb{N}^{+}}

\newcommand{\Integers}{\mathbb{Z}}
\newcommand{\Rationals}{\mathbb{Q}}
\newcommand{\Weight}{w}
\newcommand{\Dim}{d}
\newcommand{\Safe}{\mathsf{Safe}}
\newcommand{\Live}{\mathsf{Live}}
\newcommand{\Paths}{\Omega}

\newcommand{\MPI}{{\sf MP}}
\newcommand{\MP}{{\sf MP}}
\newcommand{\R}{{\sf Ratio}}
\newcommand{\Ratio}{{\sf Ratio}}
\newcommand{\In}{{\sf IN}}
\newcommand{\Out}{{\sf OUT}}
\newcommand{\Perm}{\mathsf{Perm}}

\newcommand{\Vscc}{V_{\text{SCC}}}
\newcommand{\Escc}{E_{\text{SCC}}}
\newcommand{\trans}{\Delta}

\newcommand{\A}{{\mathcal{A}}}

\newcommand{\C}{{\mathcal{C}}}
\newcommand{\NP}{{\sc NP}}
\renewcommand{\S}{{\mathcal{S}}}
\renewcommand{\L}{{\mathcal{L}}}
\newcommand{\W}{{\mathcal{W}}}
\newcommand{\CR}{{\mathcal{CR}}}

\newcommand{\ClairvoyantSuccessor}{\mathsf{ClairvoyantSuccessor}}
\newcommand{\Knapsack}{\mathsf{knapsack}}
\newcommand{\AdaptiveBinarySearch}{\mathsf{AdaptiveBinarySearch}}
\newcommand{\J}{{\mathcal{J}}}

\newcommand{\Tau}{{\mathcal{T}}}

\newcommand{\InAct}{\Sigma}
\newcommand{\OutAct}{\Pi}

\renewcommand{\geq}[0]{\geqslant}
\renewcommand{\leq}[0]{\leqslant}

\newcommand{\miover}{\mathit{over}}
\let\emptyset\varnothing

\begin{document}
\title{
A Framework for Automated Competitive Analysis of On-line Scheduling of Firm-Deadline Tasks\thanks{{This work has been supported by the Austrian Science Foundation (FWF)
under the NFN RiSE (S11405 and S11407), FWF Grant P23499-N23, 
ERC Start grant (279307: Graph Games), and Microsoft faculty fellows award.}}}

\author[1]{Krishnendu Chatterjee\thanks{krish.chat@ist.ac.at}}
\author[1]{Andreas Pavlogiannis\thanks{pavlogiannis@ist.ac.at}}
\author[2]{Alexander K\"o\ss ler\thanks{koe@ecs.tuwien.ac.at}}
\author[2]{Ulrich Schmid\thanks{s@ecs.tuwien.ac.at}}
\affil[1]{IST Austria (Institute of Science and Technology Austria)}
\affil[2]{Embedded Computing Systems Group, Vienna University of Technology, Vienna, Austria}

\date{}
\maketitle
\begin{abstract}
We present a flexible framework for the automated competitive analysis of
on-line scheduling algorithms for firm-deadline real-time tasks
based on multi-objective graphs: Given a taskset and an on-line scheduling 
algorithm specified as a labeled transition system, along with some optional 
safety, liveness, and/or limit-average constraints for the adversary, 
we automatically compute the competitive ratio of the  algorithm w.r.t.\
a clairvoyant scheduler. We demonstrate the flexibility and power of
our approach by comparing the competitive ratio of several on-line algorithms, 
including $D^{\miover}$, that have been proposed in
the past, for various tasksets. Our experimental results reveal that none of these algorithms is 
universally optimal, in the sense that there are tasksets where other schedulers
provide better performance. Our framework is hence a very useful design
tool for selecting optimal algorithms for a given application.
\end{abstract}

%
%
%

\vspace{-0.0em}
\section{Introduction}
\label{sec:intro}

We study the well-known problem of scheduling a sequence of dynamically
arriving real-time task instances with firm deadlines on a single processor 
using a novel approach, namely, automated competitive analysis based on a corresponding
multi-objective graph representation. In firm deadline scheduling, a task instance 
(a \emph{job}) that is completed by its deadline contributes a positive utility 
value; a job that does not meet its deadline
does not harm, but does not add any utility. The goal of the scheduling algorithm
is to maximize the cumulated utility. Firm deadline tasks arise in 
various application domains, e.g., machine scheduling, 
multimedia and video streaming, QoS management in switches and
data networks, and other systems that may suffer from overload~\cite{KS95}.

Competitive analysis \cite{BE98} has been the primary tool for studying
the performance of such scheduling algorithms \cite{BKMM92}. In general, it
allows to compare the performance of an \emph{on-line} 
algorithm $\A$, which processes a sequence of inputs without knowing
the future, with what can be achieved by an optimal \emph{off-line}
algorithm $\C$ that does know the future (a \emph{clairvoyant} algorithm):
The \emph{competitive factor} gives the worst-case performance ratio of $\A$ 
vs. $\C$ over all possible scenarios.

In a seminal paper \cite{BKMM92}, Baruah et~al.\ proved that no on-line 
scheduling algorithm for single processors can achieve a competitive factor better than $1/4$ over
a clairvoyant algorithm in \emph{all} possible job sequences of \emph{all} possible 
tasksets. The proof is based on constructing a specific job sequence,
which takes into account the on-line algorithm's actions 
and thereby forces any such algorithm to deliver a sub-optimal cumulated utility.
For the special case of zero-laxity tasksets of uniform value-density, where utilities equal execution times,
they also provided the on-line algorithm TD1 with competitive factor~$1/4$,
concluding that $1/4$ is a tight bound for this family of tasksets.
In~\cite{BKMM92}, the $1/4$ upper bound was also generalized, by showing
that there exist tasksets with \emph{importance ratio} $k$,
defined as the ratio of the maximum over the minimum value-density in the taskset,
in which no on-line scheduler can have competitive factor larger than $\frac{1}{(1+\sqrt{k})^2}$. In a subsequent work~\cite{KS95},
the on-line scheduler $D^{\miover}$ was introduced, which
provides the performance guarantee of $\frac{1}{(1+\sqrt{k})^2}$ in any
taskset with importance ratio $k$, showing that this bound is also tight.

Since the taskset arising in a particular application is usually known, 
our paper focuses on the competitive analysis problem for \emph{given} tasksets: Rather
than from \emph{all} possible tasksets as in \cite{BKMM92}, the job sequences used for computing
the \emph{competitive ratio} are chosen from a taskset given as an input.
There are two relevant problems for the automated competitive analysis for 
a given taskset: (1)~The \emph{synthesis} question asks to find an algorithm 
with optimal competitive ratio; and (2)~the \emph{analysis} question asks to 
compute the competitive ratio of a given on-line algorithm.
In \cite{CKS13:HSCC}, we studied the synthesis problem and presented a reduction 
to a problem in graph games~\cite{Vel12}, which we showed to be 
\NP-complete.

In this paper, we consider the analysis problem.
More specifically, we provide a flexible, automated analysis framework that 
also supports additional constraints on the adversary, such as sporadicity 
constraints and longrun-average load.
We show that the analysis problem (with additional constraints) can be reduced
to a multi-objective graph problem, which can be solved in polynomial time.
We also present several optimizations
and an experimental evaluation of 
our algorithms that demonstrates the feasibility of our approach,
which effectively allows to replace human ingenuity 
(required for finding worst-case scenarios)
by computing power: 
Using our framework,
the application designer can analyze different scheduling algorithms for the specific 
tasksets arising in her/his particular application, and compare their competitive 
ratio in order to select the best one. 

\smallskip\noindent{\em Detailed contributions and paper organization:}
\begin{compactenum}
\item In Section~\ref{sec:problemdef}, we define our scheduling problem along 
with the relevant additional constraints on the adversary. 

\item In Section~\ref{sec:lts}, we introduce the labeled transition systems as a formal model for 
specifying on-line and off-line algorithms.
In Section~\ref{sec:constraints}, we present the formal framework to specify the constraints on the adversary, 
and argue how it allows to model a wide variety of constraints.
We also give an overview of all the steps involved in our approach.

\item In Section~\ref{sec:graphs}, we present the \emph{multi-objective graphs} used 
by our solution algorithm. Multiple objectives 
are required to represent the competitive analysis problem with various constraints.

\item In Section~\ref{sec:reduction}, we describe a theoretical reduction of the
competitive analysis problem to solving a multi-objective graph problem, where the
graph is obtained as a product of the on-line algorithm, an off-line algorithm, and 
the constraints specified as automata.
Our algorithmic solution is polynomial in the size of the graph; however, the product 
graph can be large for representative tasksets.

\item In Section~\ref{sec:optimization}, we present both general and implementation-specific
optimizations, which considerably reduce the size of the resulting graphs.

\item In Section~\ref{sec:results}, we provide competitive ratio analysis results 
obtained by our method. More specifically,
we present  a comparative study of the performance of several 
existing firm deadline real-time scheduling algorithms.
Our results show that, for different tasksets (even with no constraints),
different algorithms achieve the highest competitive ratio (i.e., there is no universal optimal
algorithm).
Moreover, even for a fixed taskset and varying constraints on the adversary, 
different algorithms achieve the highest competitive ratio.
This highlights the importance of our framework for selecting optimal algorithms for specific applications.
\end{compactenum}


\smallskip\noindent{\em Related work:} 
Algorithmic game theory \cite{NRTV07} has been applied to classic 
scheduling
problems since decades, primarily in economics and operations research, see 
e.g.\ \cite{Kou11} for just one example of some more recent work. 
It has also been applied for real-time scheduling of \emph{hard} 
real-time tasks in the past: Besides Altisen et~al.\ \cite{AGS02}, who
used games for synthesizing controllers dedicated to meeting all
deadlines, Bonifaci and Marchetti-Spaccamela \cite{BM12} employed
graph games for automatic feasibility analysis
of sporadic real-time tasks in multiprocessor systems: Given
a set of sporadic tasks (where consecutive releases of jobs of 
the same task are separated at least by some sporadicity interval), the
algorithms provided in \cite{BM12} allow to decide, in polynomial time,
whether some given scheduling algorithm will meet \emph{all} deadlines.
A partial-information game variant of their approach also allows
to synthesize an optimal scheduling algorithm for a given taskset
(albeit not in polynomial time).
As these approaches do not generalize to competitive analysis of 
tasks with firm deadlines, we studied the related synthesis problem 
in~\cite{CKS13:HSCC}.

Regarding firm deadline task scheduling in general, 
starting out from \cite{BKMM92}, 
classic real-time systems research
has studied the competitive factor of both simple and 
extended real-time scheduling 
algorithms. The competitive analysis of simple algorithms 
(see Section~\ref{sec:results} for the references) has
been extended in various ways later on:
Energy consumption \cite{AMMM04,DLA10} (including dynamic voltage scaling), 
imprecise computation tasks (having both a mandatory and an
optional part and associated utilities) \cite{BH98}, lower bounds
on slack time \cite{BH97}, and fairness \cite{Pal04}. Note that
dealing with these extensions involved considerable ingenuity and efforts w.r.t.\
identifying and analyzing appropriate worst case scenarios, which
do not necessarily carry over even to minor variants of the problem.
Maximizing cumulated utility while satisfying multiple resource constraints
is also the purpose of the Q-RAM (QoS-based Resource Allocation Model) 
\cite{RLLS97} approach. 

\section{Problem Definition}
\label{sec:problemdef}

\noindent{\em Real-time scheduling setting.}
We consider a finite set of tasks $\Tau=\{\tau_1,\dots,\tau_N\}$,  to be 
executed on a single processor. 
We assume a discrete notion of real-time $t=k\varepsilon$, $k\geq 1$, 
where $\varepsilon>0$ is both the unit time and the smallest unit of 
preemption (called a \emph{slot}). Since both task releases 
and scheduling activities occur at slot boundaries only,
all timing values are specified as positive integers.
Every task $\tau_i$ releases countably
many task instances (called \emph{jobs}) 
$J_{i,j}:=(\tau_i,j)\in \Tau \times \Natsplus$ (where $\Natsplus$ is the set of 
positive integers) over 
time (i.e., $J_{i,j}$ denotes that a job of task $i$ is released at time $j$). All jobs, of all tasks, are independent 
of each other and can be preempted and resumed during execution 
without any overhead.
Every task $\tau_i$, for $1\leq i \leq N$, is characterized by a 3-tuple
$\tau_i=(C_i,D_i,V_i)$ consisting of its non-zero \emph{worst-case 
execution time} $C_i\in\Natsplus$ (slots), its non-zero \emph{relative deadline} 
$D_i\in\Natsplus$ (slots) and its non-zero \emph{utility value} 
$V_i \in \Natsplus$ (rational utility values $V_1,\dots,V_n$ 
can be mapped to integers by proper scaling).  We denote with $D_{\max}=\max_{1\leq i \leq N} D_i$ the maximum relative deadline in $\Tau$.
Every job $J_{i,j}$ needs the processor for $C_i$ (not necessarily consecutive)
slots exclusively to execute to completion. All tasks have firm deadlines: 
only a job $J_{i,j}$ that completes
within $D_i$ slots, as measured from its release time, provides utility 
$V_i$ to the system. 
A job that misses its deadline does not harm but provides zero utility.
The goal of a real-time scheduling algorithm in this model is to 
maximize the \emph{cumulated utility},
which is the sum of $V_i$ times the number of jobs $J_{i,j}$ that can 
be completed by their deadlines, in a 
sequence of job releases generated by the \emph{adversary}. 

\smallskip\noindent{\em Notation on sequences.}
Let $X$ be a finite set. 
For an infinite sequence $x=(x^\ell)_{\ell\geq 1}=(x^1,x^2, \ldots)$ of elements in $X$,
we denote by $x^{\ell}$ the element in the $\ell$-th position of $x$,
and denote by $x(\ell)=(x^1,x^2,\ldots,x^{\ell})$ the finite prefix
of $x$ up to position $\ell$. We denote by $X^\infty$ the set of 
all infinite sequences of elements from $X$.
Given a function $f: X\rightarrow \Integers$ (where $\Integers$ is the set of 
integers) and a sequence $x\in X^{\infty}$, 
we denote with $f(x, k)=\sum_{\ell=1}^kf(x^{\ell})$ the sum of the images of the first $k$ elements.

\smallskip\noindent{\em Job sequences.}
When generating a job sequence, the adversary releases at most one 
new job from every task in every slot.
Formally, the adversary generates an infinite \emph{job sequence} 
$\sigma=(\sigma^{\ell})_{\ell\geq 1}\in \Sigma^\infty$,
where $\Sigma=2^{\Tau}$. If a task $\tau_i$ belongs to $\sigma^\ell$, 
for $\ell \in \Natsplus$, then a (single) new job $J_{i,j}$ of task $i$ is 
released at the beginning of slot $\ell$: 
$j=\ell$ denotes the \emph{release time} of $J_{i,j}$, which is the earliest 
time $J_{i,j}$ can be executed, and $d_{i,j}=j+D_i$ denotes its 
absolute \emph{deadline}.

\smallskip\noindent{\em Admissible job sequences.}
We present a flexible framework where the set of admissible job sequences that 
the adversary can generate may be restricted.
The set $\J$ of \emph{admissible} job sequences from $\Sigma^\infty$ 
can be obtained by imposing one or more of the following (optional) 
admissibility restrictions: 
\begin{compactenum}
\item[($\S$)] Safety constraints, which are restrictions that hold
in every finite prefix of a job sequence; e.g., they can 
be used to enforce job release constraints such as periodicity or sporadicity, 
and to impose temporal workload restrictions.
\item[($\L$)] Liveness constraints, which assert infinite repetition of certain
patterns in a job sequence; e.g., they can be used to force the
adversary to release a certain task infinitely often.
\item[($\W$)] Limit-average constraints, which restrict the long run average behavior
of a job sequence; e.g., they can be used to enforce
that the average load in the job sequences does not exceed a threshold.
\end{compactenum}

%


\smallskip\noindent{\em Schedule.}
Given an admissible job sequence $\sigma\in \J$, the \emph{schedule} 
$\pi=(\pi^{\ell})_{\ell\geq 1}\in \OutAct^\infty$, where $\OutAct=((\Tau\times \{0,\dots,D_{\max}-1\}) \cup \emptyset)$, computed 
by a real-time scheduling algorithm for $\sigma$, is a function 
that assigns at most one job for execution
to every slot $\ell\geq 1$: $\pi^\ell$ is either $\emptyset$ 
(i.e., no job is executed) or else $(\tau_i,j)$ (i.e., the job 
$J_{i,\ell-j}$  of task $\tau_i$ released $j$ slots ago is executed). 
The latter must satisfy the following constraints: 
\begin{compactenum}
\item $\tau_i\in \sigma^{\ell-j}$ (the job has been released),

\item \label{constraint:schedbeforedeadline} $j < D_i$ (the job's deadline has not passed),
\item $|\{k:\mbox{$k>0$ and $\pi^{\ell-k}=(\tau_i,j')$ and $k+j'=j$}\}|<C_i$ (the job 
released in slot $\ell-j$ has not been completed).
\end{compactenum}
Note that our definition of schedules uses relative indexing in the scheduling algorithms: 
At time point $\ell$, the algorithm for schedule $\pi^\ell$ uses index $j$ to refer to slot $\ell-j$.
Recall that $\pi(k)$ denotes the prefix of length $k\geq 1$ of $\pi$. We define
$\gamma_i(\pi,k)$ to be the number of jobs of task $\tau_i$ that are completed 
by their deadlines in $\pi(k)$. The cumulated utility $V(\pi,k)$ (also
called utility for brevity) achieved 
in $\pi(k)$ is defined as $V(\pi,k)=\sum_{i=1}^N \gamma_i(\pi,k)\cdot V_i$.

\smallskip\noindent{\em Competitive ratio.}
We are interested in evaluating the performance of
deterministic \emph{on-line} scheduling algorithms $\A$, which, 
at time $\ell$, do not know any of the $\sigma^k$ for $k>\ell$
when running on $\sigma\in\J$. 
In order to assess the performance of $\A$, we will
compare the cumulated utility achieved in the schedule $\pi_{\A}$ to the 
cumulated utility achieved in the schedule $\pi_{\C}$ provided
by an optimal \emph{off-line} scheduling algorithm, called a \emph{clairvoyant} algorithm $\C$, 
working on the same job sequence. Formally, 
given a taskset $\Tau$, let $\J\subseteq \Sigma^{\infty}$ be the set of all 
admissible job sequences of $\Tau$ that satisfy given (optional) safety, 
liveness, and limit-average constraints. 
For every $\sigma\in \J$, we denote with $\pi_{\A}^\sigma$ 
(resp.\ $\pi_{\C}^\sigma$) the schedule produced by $\A$ (resp.\ $\C$) under 
$\sigma$. 
The \emph{competitive ratio} of the on-line algorithm $\A$ for the taskset 
$\Tau$ under the admissible job sequence set $\J$ is defined as
\begin{equation}
\CR_{\J}(\A)=\inf_{\sigma\in\J}\liminf_{k\to\infty} \frac{1+V(\pi_{\A}^\sigma,k)}{1+V(\pi_{\C}^\sigma,k)}\label{equ:CR}
\end{equation}
that is, the worst-case ratio of the cumulated utility of the on-line algorithm versus the clairvoyant algorithm, 
under all admissible job sequences.
Note that adding~1 in numerator and denominator simply avoids division by zero issues.

\begin{remark}
Since, according to the definition of the competitive ratio $\CR_{\J}$
in Equation~(\ref{equ:CR}), we focus on worst-case analysis, we do not consider 
randomized algorithms (such as Locke's best-effort policy~\cite{Locke86}).
Generally, for worst-case analysis, randomization can be handled by 
additional choices for the adversary. 
For the same reason, we do not consider scheduling algorithms that can use the 
unbounded history of job releases to predict the future 
(e.g., to capture correlations).
\end{remark}

\section{LTSs as Models for Algorithms}\label{sec:lts}

We will consider both on-line and off-line scheduling algorithms
that are formally modeled as \emph{labeled transition systems (LTSs)}: 
Every deterministic finite-state on-line scheduling algorithm can be 
represented as a deterministic LTS, such that every input job 
sequence generates a unique run that determines the corresponding
schedule. On the other hand,
an off-line algorithm can be represented as a non-deterministic LTS, which
uses the non-determinism to guess the appropriate job to schedule.

\smallskip\noindent{\em Labeled transition systems (LTSs).}
Formally, a \emph{labeled transition system } (LTS) is a tuple $L=(S, s_1, \InAct, \OutAct, \trans)$, 
where $S$ is a finite set of states, $s_1\in S$ is the initial state, $\InAct$ is a finite set of input actions, 
$\OutAct$ is a finite set of output actions, and $\trans \subseteq S\times\InAct\times S\times \OutAct$ is the transition relation. 
Intuitively, $(s,x,s',y)\in \trans$ if, given the current state $s$ and input $x$, the LTS outputs $y$ and makes a transition to state $s'$. 
If the LTS is deterministic, then there is always a unique output 
and next state, i.e., $\trans$ is a function $\trans:  S\times\InAct \to S\times\OutAct$.
Given an input sequence $\sigma\in \InAct^{\infty}$, a \emph{run} of $L$ on $\sigma$ is a sequence $\rho=(p_{\ell},\sigma_{\ell}, q_{\ell}, \pi_{\ell})_{\ell\geq 1}\in \trans^{\infty}$ such that $p_1=s_1$ and for all $\ell\geq 2$, we have $p_{\ell}=q_{\ell-1}$. 
For a deterministic LTS, for each input sequence, there is a unique run.

\smallskip\noindent{\em Deterministic LTS for an on-line algorithm.}
For our analysis, on-line scheduling algorithms are represented as deterministic LTSs.
Recall the definition of the sets $\Sigma=2^{\Tau}$, and  $\OutAct=((\Tau\times \{0,\dots,D_{\max}-1\}) \cup \emptyset)$.
Every deterministic on-line algorithm $\A$ that uses finite state space (for all job sequences) 
can be represented as a deterministic LTS $L_{\A}=(S_{\A},s_{\A},\InAct, \OutAct,\trans_{\A})$,
where the states $S_{\A}$ correspond to the state space of $\A$, and $\trans_{\A}$ correspond 
to the execution of $\A$ for one slot. 
Note that, due to relative indexing, for every current slot $\ell$, the schedule $\pi^{\ell}$ of $\A$ contains elements 
from the set $\Pi$, and $(\tau_i,j)\in \pi^{\ell}$ uniquely determines the job $J_{i,\ell-j}$. 
Finally, we associate with $L_{\A}$ a reward function $r_{\A}:\Delta_{\A}\rightarrow \Nats$ such that 
$r_{\A}(\delta)=V_i$ if the transition $\delta$ completes a job of task $\tau_i$, and $r_{\A}(\delta)=0$ otherwise.
Given the unique run $\rho^\sigma=(\delta^\ell)_{\ell\geq1}$ of $L_{\A}$
for the job sequence $\sigma$, where
$\delta^\ell$ denotes the transition taken at the beginning of slot $\ell$,
the cumulated utility in the prefix of the first $k$
transitions in $\rho^\sigma$ is $V(\rho^\sigma,k)=\sum_{\ell=1}^k r_{\A}(\delta^{\ell})$.

Most scheduling algorithms (such as EDF, FIFO, DOVER, TD1) 
can be represented as a deterministic LTS.
An illustration for EDF is given in the following example (see Appendix Section~\ref{sec:ex-sch} for other examples).

\begin{figure}
\centering
\includegraphics[scale=0.18]{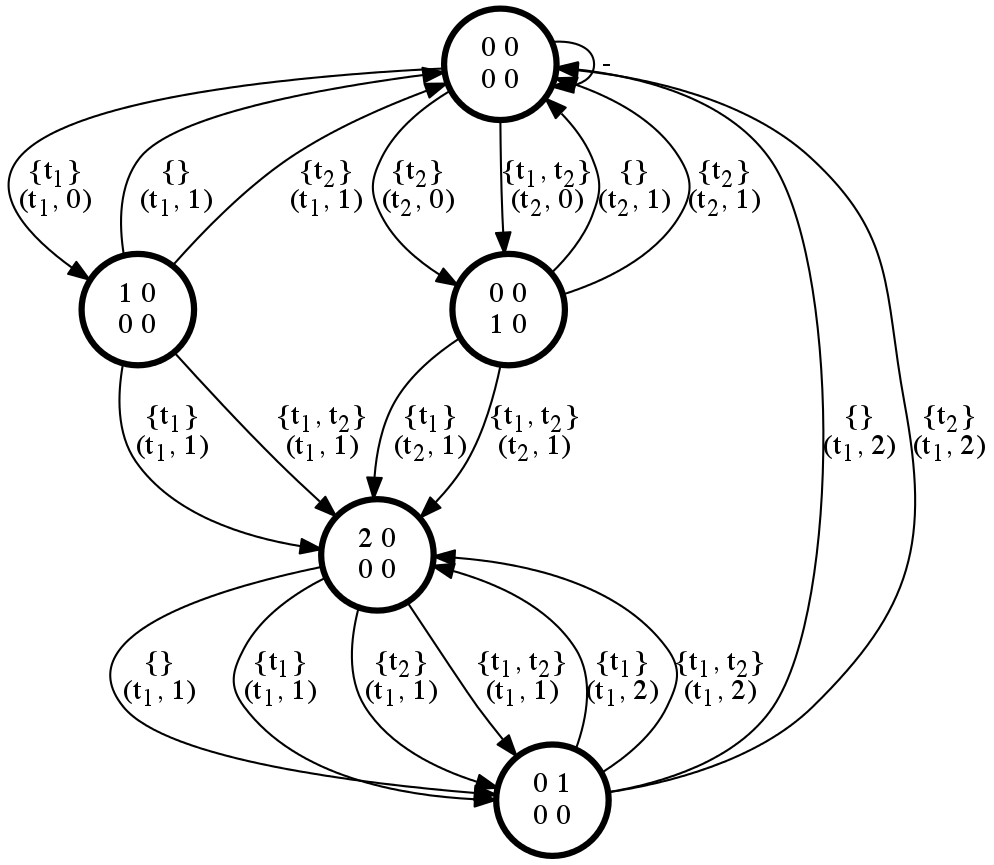}
\caption{EDF for $\Tau=\{\tau_1, \tau_2\}$ with $D_1=3$, $D_2=2$ and 
$C_1=C_2=2$, represented as a deterministic LTS.}\label{fig:edf_set1}
\end{figure}

\begin{example}\label{ex:edf}
Consider the taskset $\Tau=\{\tau_1, \tau_2\}$, with $D_1=3$, $D_2=2$ and $C_1=C_2=2$.
Figure~\ref{fig:edf_set1} represents the EDF (Earliest Deadline First) scheduling policy as a deterministic LTS for $\Tau$. 
Each state is represented by a matrix $M$, such that $M[i,j]$, $1\leq i \leq N$,
$1\leq j \leq D_{\max}-1$, denotes the remaining execution time
of the job of task $\tau_i$ released $j$ slots ago. 
Every transition is labeled with a set $T\in \Sigma$ of released tasks as well
as with $(\tau_i, j)\in \Pi$, which denotes the unique job $J_{i,\ell-j}$ to be scheduled in the current slot~$\ell$. 
Released jobs with no chance of being scheduled 
are not included in the state space.
\end{example}

\smallskip\noindent{\em The non-deterministic LTS.}
The clairvoyant algorithm $\C$ is formally a non-deterministic LTS $L_{\C}=(S_{\C},s_{\C},\InAct, \OutAct,\trans_{\C})$ where each state in $S_{\C}$ is a $N\times (D_{\max}-1)$ matrix $M$, such that for each time slot $\ell$, the entry $M[i,j]$, $1\leq i \leq N$,
$1\leq j \leq D_{\max}-1$, denotes the remaining execution time
of the job $J_{i,\ell-j}$ (i.e., the job of task $i$ released $j$ slots ago). 
For matrices $M$, $M'$, subset $T\in \InAct$ of newly released tasks, 
and scheduled job $P=(\tau_i,j)\in \OutAct$, we have $(M,T,M',P)\in \trans_{\C}$ iff $M[i,j]>0$ and $M'$ is obtained from $M$ by 
\begin{compactenum}
\item[(1)] inserting all $\tau_i\in T$ into $M$, 
\item[(2)] decrementing the value at position $M[i,j]$, and 
\item[(3)] shifting the contents of $M$ by one column to the right. 
\end{compactenum}
That is, $M'$ corresponds to $M$ after inserting all released tasks in the current state, executing a pending task for one unit of time, and reducing the relative deadlines of all tasks currently in the system. The initial state $s_{\C}$ is represented by the zero $N\times (D_{\max}-1)$ 
matrix, and $S_{\C}$ is the smallest $\trans_{\C}$-closed set of states that contains $s_{\C}$ (i.e., if $M\in S_{\C}$ and $(M,T,M',P)\in \trans_{\C}$ for some $T$, $M'$ and $P$, we have $M'\in S_{\C}$). Finally, we associate with $L_{\C}$ a reward function $r_{\C}:\Delta_{\C}\rightarrow \Nats$ such that $r_{\C}(\delta)=V_i$ if the transition $\delta$ completes a task $\tau_i$, and $r_{\C}(\delta)=0$ otherwise.

\section{Admissible Job Sequences and Our Approach}\label{sec:constraints}
In this section we discuss our mechanisms for restricting the adversary 
to generate only certain admissible job sequences and then present our 
overall approach.

\smallskip\noindent{\em Admissible job sequences.}
Our framework allows to restrict the adversary to generate
admissible job sequences $\J\subseteq \Sigma^{\infty}$, 
which can be specified via
different constraints. Since a constraint on job sequences 
can be interpreted as a language (which is a subset
of infinite words $\Sigma^\infty$ here), we will use automata as acceptors of 
such languages. Since an automaton is a deterministic LTS with no output,
all our constraints will be described as LTSs with an empty set of output 
actions. We allow the following types of contraints:
\begin{compactenum}
\item[($\S$)] Safety constraints are defined by a deterministic safety LTS $L_{\S}=(S_{\S}, s_{\S}, \InAct, \emptyset, \trans_{\S})$, 
with a distinguished \emph{absorbing} reject state $s_{r}\in S_{\S}$. 
An absorbing state is a state that has outgoing transitions only to itself.
Every job sequence $\sigma$ defines a unique run $\rho_{\S}^{\sigma}$ in $L_{\S}$, such that either no transition to $s_{r}$ 
appears in $\rho_{\S}^{\sigma}$, or every such transition is followed solely 
by self-transitions to $s_r$. 
A job sequence $\sigma$ is \emph{admissible} to $L_{\S}$, if $\rho_{\S}^{\sigma}$ does not contain a transition to $s_r$. To obtain a safety LTS that does 
not restrict $\J$ at all, we simply use a trivial deterministic $L_{\S}$ with 
no transition to $s_r$.

\item[($\L$)] Liveness constraints are defined by a deterministic liveness LTS $L_{\L}=(S_{\L}, s_{\L}, \InAct, \emptyset, \trans_{\L})$, 
with a distinguished \emph{accept} state $s_a\in S_{\L}$. A job sequence $\sigma$ is \emph{admissible} to $L_{\L}$ if $\rho_{\L}^{\sigma}$ 
contains infinitely many transitions to $s_a$.
For the case where there are no liveness constraint in $\J$, 
we use a LTS $L_{\L}$ consisting of state $s_a$ only.
 
\item[($\W$)] Limit-average constraints are defined by a deterministic weighted LTS 
$L_{\W}=(S_{\W}, s_{\W}, \InAct, \emptyset, \trans_{\W})$ equipped with a weight function $\Weight:\trans_{\W}\rightarrow \Integers^{\Dim}$ 
that assigns a vector of weights to every transition. 
Given a threshold vector $\vec{\lambda}\in \Rationals^{\Dim}$, where $\Rationals$
denotes the set of all rational numbers, 
a job sequence $\sigma$ and the corresponding run 
$\rho_{\W}^{\sigma}$ of $L_{\W}$, the job sequence is \emph{admissible} to $L_{\W}$ 
if $\liminf_{k\to\infty} \frac{1}{k}\cdot\Weight(\rho_{\W}^\sigma,k)\leq \vec{\lambda}$.
\end{compactenum}

\smallskip\noindent{\em Illustrations of admissible job sequences.}
We now illustrate the types of constraints that are supported
by the above framework with some examples.
\begin{compactenum}
\item[($\S$)] {\em Safety constraints.}
Safety constraints restrict the adversary to release job sequences, where
every finite prefix satisfies some property (as they lead to the absorbing
reject state $s_r$  of $L_{\S}$ otherwise). 
Some well-known examples of safety constraints are (i) periodicity and/or sporadicity 
constraints, where there are fixed and/or a minumum time between 
the release of any two consecutive jobs of a given task,
and (ii)~absolute workload constraints~\cite{Gol91,Cru91a}, 
where the total workload released in the last $k$ slots, for some fixed $k$, 
is not allowed to exceed a threshold $\lambda$. 
For example, in case of absolute 
workload constraints, $L_{\S}$ simply encodes the workload in 
the last $k$ slots in its state, and makes a transition to $s_r$ 
whenever the workload exceeds $\lambda$.


\item[($\L$)] {\em Liveness constraints.}
Liveness constraints force the adversary to release job sequences that 
satisfy some property infinitely often. For example, they could be used
to guarantee that the release of some particular task $\tau_i$ does 
not eventually
stall; the constraint is specified by a two-state LTS $L_{\L}$ that visits $s_a$ whenever the 
current job set includes $\tau_i$. A liveness
constraint can  also be used to prohibit infinitely long periods of overload~\cite{BKMM92}.

\item[($\W$)] {\em Limit-average constraints.} 
Consider a relaxed notion of workload constraints, where the adversary is restricted to generate job 
sequences whose \emph{average} workload does not exceed a threshold $\lambda$.
Since this constraint still allows ``busy'' intervals where the workload temporarily exceeds $\lambda$, 
it cannot be expressed as a safety constraint. To support such interesting average constraints of 
admissible job sequences, where the adversary is more relaxed than under
absolute constraints,
our framework explicitly supports limit-average constraints.
Therefore, it is possible to express the average workload assumptions 
commonly used in the analysis of aperiodic task scheduling in soft-real time 
systems \cite{AB98,HCL90}.
Other interesting cases of limit-average constraints include restricting the 
average sporadicity,  and, in particular, average energy: 
ensuring that the limit-average
of the energy consumption is below a certain threshold is an important
concern
in modern real-time systems~\cite{AMMM04}.
\end{compactenum}

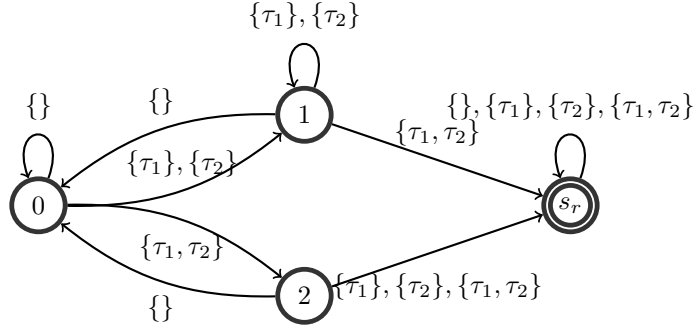
\begin{figure}[!h]
\centering
\begin{tikzpicture}[thick,
pre/.style={<-,shorten >= 1pt, shorten <=1pt, thick},
post/.style={->,shorten >= 1pt, shorten <=1pt,  thick},
und/.style={very thick, draw=gray},
bag/.style={ellipse, minimum height=7mm,minimum width=14mm,draw=gray!80, line width=1pt},
rootbag/.style={ellipse, minimum height=7mm,minimum width=14mm,draw=black!80, line width=2.5pt},
node/.style={circle,draw=black!80, inner sep=1, minimum size=20pt, line width=1.8},
virt/.style={circle,draw=black!50,fill=black!20, opacity=0}]

\node	[node]		(q1)		at	(0,0)	{$0$};
\node	[node]		(q2)		at	(3.5,1.2)	{$1$};
\node	[node]		(q3)		at	(3.5,-1.2)	{$2$};
\node	[node]		(q4)		at	(7,0)	{};
\node	[node, minimum size=15pt]		(accept)	at	(7,0)	{$s_r$};

\draw [->, thick, loop below, looseness=8, out=90-20, in=90+20] (q1) to  node [auto, above=2] {$\{\}$} 	(q1);
\draw [->, thick, loop below, looseness=8, out=90-20, in=90+20] (q2) to  node [auto, above=2] {$\{\tau_1\}, \{\tau_2\}$} 	(q2);
\draw [->, thick, loop below, looseness=8, out=90-20, in=90+20] (q4) to  node [auto, above=2] {$\{\},\{\tau_1\}, \{\tau_2\},\{\tau_1,\tau_2\}$} 	(q4);

\draw [->, thick, bend right=20] (q1) to  node [auto, minimum size=2pt, inner sep=5, above=0] {$\{\tau_1\}, \{\tau_2\}$} 	(q2);
\draw [->, thick, bend left=20] (q1) to  node [auto, minimum size=2pt, inner sep=5, below=0] {$\{\tau_1, \tau_2\}$} 	(q3);
\draw [->, thick, bend right=20] (q2) to  node [auto, minimum size=2pt, inner sep=5, above=0] {$\{\}$} 	(q1);
\draw [->, thick] (q2) to  node [auto, minimum size=2pt, inner sep=5, above=0] {$\{\tau_1,\tau_2\}$} 	(q4);
\draw [->, thick, bend left=20] (q3) to  node [auto, minimum size=2pt, inner sep=5, below=0] {$\{\}$} 	(q1);
\draw [->, thick] (q3) to  node [auto, minimum size=2pt, inner sep=8, below=0] {$\{\tau_1\},\{\tau_2\},\{\tau_1, \tau_2\}$} 	(q4);

\end{tikzpicture}
\caption{Example of a safety LTS $L_{\S}$ that restricts the adversary to at most $2$ units of workload in the last $2$ rounds.}\label{fig:lts_safe}
\end{figure}

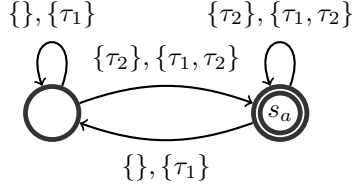
\begin{figure}[!h]
\centering
\begin{tikzpicture}[thick,
pre/.style={<-,shorten >= 1pt, shorten <=1pt, thick},
post/.style={->,shorten >= 1pt, shorten <=1pt,  thick},
und/.style={very thick, draw=gray},
bag/.style={ellipse, minimum height=7mm,minimum width=14mm,draw=gray!80, line width=1pt},
rootbag/.style={ellipse, minimum height=7mm,minimum width=14mm,draw=black!80, line width=2.5pt},
node/.style={circle,draw=black!80, inner sep=1, minimum size=20pt, line width=1.8},
virt/.style={circle,draw=black!50,fill=black!20, opacity=0}]

\node	[node]		(q1)		at	(0,0)	{};
\node	[node]		(q2)		at	(3,0)	{};
\node	[node, minimum size=15pt]		(accept)	at	(3,0)	{$s_a$};

\draw [->, thick, loop below, looseness=8, out=90-20, in=90+20] (q1) to  node [auto, above=2] {$\{\}, \{\tau_1\}$} 	(q1);
\draw [->, thick, loop below, looseness=8, out=90-20, in=90+20] (q2) to  node [auto, above=2] {$\{\tau_2\}, \{\tau_1,\tau_2\}$} 	(q2);

\draw [->, thick, bend left=20] (q1) to  node [auto, minimum size=2pt, inner sep=5, above=0] {$\{\tau_2\}, \{\tau_1,\tau_2\}$} 	(q2);

\draw [->, thick, bend left=20] (q2) to  node [auto, minimum size=2pt, inner sep=5, below=0] {$\{\}, \{\tau_1\}$} 	(q1);

\end{tikzpicture}
\caption{Example of a liveness LTS $L_{\L}$ that forces $\tau_2$ to be released infinitely often.}\label{fig:lts_live}
\end{figure}

\begin{figure}[!h]
\centering
\begin{tikzpicture}[thick,
pre/.style={<-,shorten >= 1pt, shorten <=1pt, thick},
post/.style={->,shorten >= 1pt, shorten <=1pt,  thick},
und/.style={very thick, draw=gray},
bag/.style={ellipse, minimum height=7mm,minimum width=14mm,draw=gray!80, line width=1pt},
rootbag/.style={ellipse, minimum height=7mm,minimum width=14mm,draw=black!80, line width=2.5pt},
node/.style={circle,draw=black!80, inner sep=1, minimum size=20pt, line width=1.8},
virt/.style={circle,draw=black!50,fill=black!20, opacity=0}]

\node	[node]		(q1)		at	(0,0)	{};

\draw [->, thick, loop below, looseness=8, out=90-20, in=90+20] (q1) to  node [auto, above] {$\{\}, \Weight=0$} 	(q1);
\draw [->, thick, loop below, looseness=8, out=270-20, in=270+20] (q1) to  node [auto, below] {$\{\tau_1, \tau_2\}, \Weight=2$} 	(q1);
\draw [->, thick, loop below, looseness=8, out=0-20, in=0+20] (q1) to  node [auto, right] {$\{\tau_1\}, \Weight=1$} 	(q1);
\draw [->, thick, loop below, looseness=8, out=180-20, in=180+20] (q1) to  node [auto, left] {$\{\tau_2\}, \Weight=1$} 	(q1);

\end{tikzpicture}
\caption{Example of a limit-average LTS $L_{\W}$ that tracks the average workload of jobs released by the adversary.}\label{fig:lts_avg}
\end{figure}
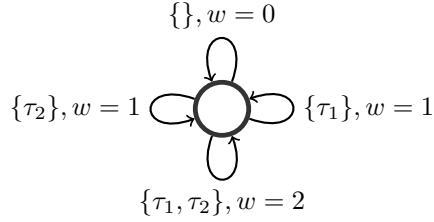

Figures~\ref{fig:lts_safe},~\ref{fig:lts_live} and ~\ref{fig:lts_avg} show examples of constraint LTSs for a taskset $\Tau=\{\tau_1, \tau_2\}$ with $C_1=C_2=1$.

\begin{remark}
While in general constraints are encoded as independent automata, 
it is often possible to encode certain constraints directly in the 
non-deterministic LTS of the clairvoyant scheduler instead. In particular, this is true
when restricting the limit-average workload, generating finite intervals of overload, 
and releasing a particular job infinitely often. 
\end{remark}

\smallskip\noindent{\em Synchronous product of LTSs.}
We present the formal definition of synchronous product of two LTSs.
We consider two LTSs $L_{1}=(S_{1},s_{1},\InAct, \OutAct,\trans_{1})$ 
and $L_{2}=(S_{2},s_{2},\InAct, \OutAct,\trans_{2})$. 
The \emph{synchronous product} of $L_1$ and $L_2$ is an LTS 
$L=(S,s,\InAct, \OutAct',\trans)$ such that:
\begin{enumerate}
\item $S\subseteq S_1\times S_2$,
\item $s=(s_1, s_2)$,
\item $\OutAct' = \OutAct\times \OutAct$, and
\item $\Delta \subseteq S\times \InAct\times S\times \OutAct'$ such that 
$((q_1,q_2),T, (q'_1,q'_2), (P_1, P_2)) \in \Delta$ iff $(q_1, T, q'_1, P_1)\in \Delta_1$ 
and $(q_2, T, q'_2, P_2)\in \Delta_2$.
\end{enumerate}
The set of states $S$ is the smallest $\Delta$-closed subset of $S_1\times S_2$ 
that contains $s$ (i.e., $s\in S$, and for each $q\in S$, if there exist 
$q'\in S_1\times S_2$, $T\in \InAct$ and $P\in \OutAct'$ such that 
$(q, T, q', P)\in \Delta$, then $q'\in S$). That is, the synchronous product
of $L_1$ with $L_2$ captures the joint behavior of $L_1$ and $L_2$ in
every input sequence $\sigma\in \InAct^{\infty}$ 
($L_1$ and $L_2$ synchronize on input actions).
Note that if both $L_1$ and $L_2$ are deterministic, so is there synchronous 
product.
The synchronous product of $k>2$ LTSs $L_1,\dots,L_k$ is defined iteratively
as the synchronous product of $L_1$ with the synchronous product of 
$L_2,\dots,L_{k}$.

\smallskip\noindent{\em Overall approach for computing $\CR$.}
Our goal is to determine the worst-case competitive ratio
$\CR_\J(\A)$ for a given on-line algorithm $\A$. 
The inputs to the problem are the given taskset $\Tau$, an 
on-line algorithm $\A$ specified as a 
deterministic LTS $L_{\A}$, and the safety, liveness, and limit-average 
constraints specified as deterministic LTSs $L_{\S}, L_{\L}$ and $L_{\W}$, 
respectively, which constrain the admissible job sequences $\J$. 
Our approach uses a reduction to a multi-objective graph problem,
which consists of the following steps:
\begin{compactenum}

\item Construct a non-deterministic LTS $L_{\C}$ 
corresponding to the clairvoyant off-line algorithm $\C$. 
Note that since $L_{\C}$ is non-deterministic, for every admissible job sequence $\sigma$,
there are many possible runs in $L_{\C}$, of course also including the runs with maximum 
cumulative utility.


\item Take the synchronous product LTS $L_{\A} \times L_{\C} \times L_{\S}\times L_{\L}\times L_{\W}$. 
By doing so, a path in the product graph corresponds to \emph{identically} 
labeled paths in LTSs, and thus ensures that they agree on the same job sequence $\sigma$. 
This product can be represented by a multi-objective graph (see Section~\ref{sec:graphs}).

\item Employ several optimizations in order to reduce the 
size of product graph (see Section~\ref{sec:reduction} and~\ref{sec:optimization}).

\item Determine $\CR_\J(\A)$ by reducing the computation of the
ratio given in Equation~(\ref{equ:CR}) to solving a multi-objective problem 
on the product graph.

\end{compactenum}

\section{Graphs with Multiple Objectives} \label{sec:graphs}
In this section, we define various objectives on graphs and outline the 
algorithms to solve them. We later show how the competitive analysis of 
on-line schedulers reduces to the solution of this section.

\smallskip\noindent{\em Multi-graphs.}
A \emph{multi-graph} $G=(V,E)$, hereinafter called simply a \emph{graph}, 
consists of a finite set $V$ of $n$ \emph{nodes}, and a finite set of $m$ 
\emph{directed multiple edges} $E \subset V\times V\times \Natsplus$. 
For brevity, we will refer to an edge $(u,v,i)$ as $(u,v)$, when $i$ is not
relevant.  
We consider graphs in which for all $u\in V$, we have $(u,v)\in E$ for some 
$v\in V$, i.e., every node has at least one outgoing edge. 
An \emph{infinite path} $\rho$ of $G$ is an infinite sequence of edges 
$e^1, e^2,\dots $ such that for all $i\geq 1$ with $e^i=(u^i, v^i)$, we have 
$v^i=u^{i+1}$. 
Every such path $\rho$ induces a sequence of nodes $(u^i)_{i\geq 1}$, which we 
will also call a path, when the distinction is clear from the context, 
and $\rho^i$ refers to $u^i$ instead of $e^i$. Finally, we denote 
with $\Paths$ the set of all paths of $G$.

\smallskip\noindent{\em Objectives.} Given a graph $G$, an objective $\Phi$ 
is a subset of $\Paths$ that defines the desired set of paths. 
We will consider safety, liveness, mean-payoff (limit-average), and ratio 
objectives, and their conjunction for multiple objectives.

\smallskip\noindent{\em Safety and liveness objectives.} 
We consider safety and liveness objectives, both defined with respect to some 
subset of nodes $X, Y \subseteq V$. 
Given $X\subseteq V$, the \emph{safety} objective defined as 
$\Safe(X)=\{\rho\in \Paths:~\forall i\geq 1, \rho^i\not \in X \}$, 
represents the set of all paths that never visit the set $X$.
The \emph{liveness} objective  defined as $\Live(Y)=\{\rho\in\Omega:~\forall j\exists i>j 
\text { s.t. }\rho^i\in Y\}$ represents the set of all paths that 
visit $Y$ infinitely often.

\smallskip\noindent{\em Mean-payoff and ratio objectives.} 
We consider the mean-payoff and ratio objectives, defined with respect to a 
weight function and a threshold. 
A \emph{weight function}  $\Weight: E\rightarrow \Integers^{\Dim}$ assigns to 
each edge of $G$ a vector of $\Dim$ integers. 
A weight function naturally extends to paths, with $\Weight(\rho, k)=\sum_{i=1}^k \Weight(\rho^i)$. 
The \emph{mean-payoff} of a path $\rho$ is defined as:
\[
\MP(\Weight,\rho)= \liminf_{k \to \infty} \frac{1}{k}\cdot \Weight(\rho, k);
\]
i.e., it is the long-run average of the weights of the path.
Given a weight function $\Weight$ and a threshold vector 
$\vec{\nu}\in \Rationals^{\Dim}$, the corresponding objective is given as:
\[
\MPI(\Weight, \vec{\nu})=\{\rho\in \Omega:~\MP(\Weight,\rho)\leq \vec{\nu}\};
\]
that is, the set of all paths such that the mean-payoff (or limit-average) of 
their weights is at most $\vec{\nu}$ (where we consider pointwise comparision 
for vectors). 
For weight functions $\Weight_1$, $\Weight_2:E\rightarrow \Nats^{\Dim}$, 
the \emph{ratio} of a path $\rho$ is defined as:
\[
\R(\Weight_1,\Weight_2,\rho)= \liminf_{k \to \infty} \frac{\vec{\mathbf{1}}+\Weight_1(\rho,k)}{\vec{\mathbf{1}}+\Weight_2(\rho,k)},
\]
which denotes the limit infimum of the coordinate-wise ratio of the sum of weights 
of the two functions; $\vec{\mathbf{1}}$ denotes the $d$-dimensional all-1 vector.
Given weight functions $\Weight_1$, $\Weight_2$ and a threshold vector 
$\vec{\nu}\in \Rationals^{\Dim}$, the ratio objective is given as:
\[
\Ratio(\Weight_1, \Weight_2, \vec{\nu})=\{\rho\in \Omega:~\R(\Weight_1, \Weight_2,\rho)\leq \vec{\nu}\}
\]
that is, the set of all paths such that the ratio of cumulative rewards w.r.t $\Weight_1$ and 
$\Weight_2$ is at most $\vec{\nu}$. 

\begin{figure}[!h]
\centering
\begin{tikzpicture}[thick,
pre/.style={<-,shorten >= 1pt, shorten <=1pt, thick},
post/.style={->,shorten >= 1pt, shorten <=1pt,  thick},
und/.style={very thick, draw=gray},
bag/.style={ellipse, minimum height=7mm,minimum width=14mm,draw=gray!80, line width=1pt},
rootbag/.style={ellipse, minimum height=7mm,minimum width=14mm,draw=black!80, line width=2.5pt},
node/.style={circle,draw=black!80, inner sep=1, minimum size=20pt, line width=1.8},
virt/.style={circle,draw=black!50,fill=black!20, opacity=0}]

\newcommand{\bend}{40}
\newcommand{\bendtwo}{30}

\node	[node]		(x1)		at	(-2.5,-1.5)		{$1$};
\node	[node]		(x2)		at	(-0.5, -1.5)		{$2$};
\node	[node]		(x3)		at	(1.5,-1.5)		{$3$};
\node	[node]		(x4)		at	(3.25,-3)		{$4$};
\node	[node]		(x5)		at	(5,-1.5)		{$5$};

\draw [->, thick, bend right=0] (-2.5, -0.7) to  node [auto, minimum size=2pt, inner sep=5, above=0.4] {} 	(x1);
\draw [->, thick, bend left=\bend] (x1) to  node [auto, minimum size=2pt, inner sep=5, above=0] {$-1,3$} 	(x2);
\draw [->, thick, bend left=\bend] (x2) to  node [auto, minimum size=2pt, inner sep=5, below=0] {$-1,-1$} 	(x1);
\draw [->, thick, bend left=\bend] (x2) to  node [auto, minimum size=2pt, inner sep=5, above=0] {$7,7$} 	(x3);
\draw [->, thick, bend left=\bend] (x3) to  node [auto, minimum size=2pt, inner sep=5, below=0] {$6,6$} 	(x2);
\draw [->, thick, bend left=\bend] (x3) to  node [auto, minimum size=2pt, inner sep=5, above=0] {$0,-1$} 	(x5);
\draw [->, thick, bend left=0] (x3) to  node [auto, minimum size=2pt, above=0] {$-5,0$} 	(x5);
\draw [->, thick, bend left=\bend] (x5) to  node [auto, minimum size=2pt, inner sep=5, above=0] {$1,0$} 	(x3);
\draw [->, thick, bend right=\bendtwo] (x3) to  node [auto, minimum size=2pt, inner sep=5, left=0.5] {$9,9$} 	(x4);
\draw [->, thick, bend right=\bendtwo] (x4) to  node [auto, minimum size=2pt, inner sep=5,right=0.5] {$8,8$} 	(x5);
\draw [->, thick, loop below, looseness=8, out=300, in=240] (x2) to  node [auto, below] {$2,1$} 	(x2);
\end{tikzpicture}
\caption{An example of a multi-graph $G$.}\label{fig:obj1}
\end{figure}
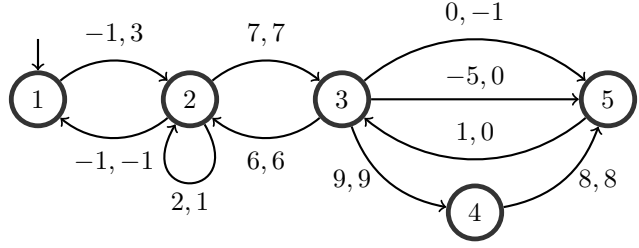

\begin{example}
Consider the multi-graph shown in Figure~\ref{fig:obj1}
with a weight function of dimension $d=2$.
Note that there are two edges from node~3 to node~5 
(represented as edges $(3,5,1)$ and $(3,5,2)$).
In the graph we have a weight function with dimension~2.
Note that the two edges from node~3 to node~5 have incomparable
weight vectors.
\end{example}

\smallskip\noindent{\em Decision problem.} 
The decision problem we consider is as follows: 
Given the graph $G$, an initial node $s\in V$, and an objective $\Phi$ (which can be a conjunction of several objectives),
determine if there exists a path $\rho$ that starts from $s$ and belongs to $\Phi$, i.e., $\rho \in \Phi$.  
For simplicity of presentation, we assume that every $u\in V$ is reachable from $s$ 
(unreachable nodes can be discarded by preprocessing $G$ in $O(m)$ time). 
We first present algorithms for each of safety, liveness, mean-payoff, and ratio obejctives separately, and 
then for their conjunction.

\smallskip\noindent{\em Algorithms for safety and liveness objectives.} 
\begin{compactenum}
\item \emph{(Safety objectives).} 
The algorithm for the objective $\Safe(X)$ is straightforward. 
We first remove the set $X$ of nodes, and iteratively remove nodes without outgoing edges. 
In the end, we obtain a graph $G=(V_X,E_X)$ such that $X \cap V_X=\emptyset$, and every node in $V_X$ has an edge to a node in $V_X$.
Thus, in the resulting graph, the objective $\Safe(X)$ is satisfied, and the algorithm answers yes iff $s \in V_X$. The algorithm requires $O(m)$ time.

\item \emph{(Liveness objectives).} 
To solve for the objective $\Live(Y)$, initially perform an SCC (maximal strongly connected component) 
decomposition of $G$. 
We call an SCC $\Vscc$ \emph{live}, if (i)~either $|\Vscc|>1$, or $\Vscc=\{u\}$ and $(u,u)\in E$; and (ii)~$\Vscc \cap Y \neq \emptyset$. 
Then $\Live(Y)$ is satisfied in $G$ iff there exists a live SCC $\Vscc$ that is reachable from $s$ 
(since every node in a live SCC can be visited infinitely often).
Using for example the algorithm of~\cite{Tar72} for the SCC decomposition also requires $O(m)$ time.
\end{compactenum}

\smallskip\noindent{\em Algorithms for mean-payoff objectives.} 
We distinguish between the case when the weight function has a single
dimension ($d=1$) versus the case when the weight function has multiple dimensions
($d > 1$).

\begin{compactenum}
\item \emph{(Single dimension).}
In the case of a single-dimensional weight function, 
a single weight is assigned to every edge, and the decision problem of the mean-payoff objective 
reduces to determining the mean weight of a minimum-weight simple cycle in $G$, as the latter
also determines the mean-weight by infinite repetition.  
Using the algorithms of~\cite{Karp78,Mad02}, this process requires $O(n\cdot m)$ time. 
When the objective is satisfied, the process also returns a simple cycle $C$, as a witness to the objective. 
From $C$, a path $\rho\in \MPI(\Weight, \vec{\nu})$ is constructed by infinite repetitions of $C$.

\item \emph{(Multiple dimensions).}
When $\Dim>1$, the mean-payoff objective reduces to determining the 
feasibility of a linear program (LP). 
For $u\in V$, let $\In(u)$ be the set of incoming, and $\Out(u)$ the set of 
outgoing edges of $u$. As shown in~\cite{Chatterjee10,Vel12}, $G$ satisfies 
$\MPI(\Weight, \vec{\nu})$ iff the following set of constraints on 
$\vec{x}=(x_e)_{e\in \Escc}$ with $x_e \in \Rationals$ is satisfied 
simultaneously on some SCC $\Vscc$ of $G$ with induced edges $\Escc\subseteq E$.

\begin{align}
& x_e\geq 0 &\ &e\in \Escc \nonumber\\
& \sum_{e\in \In(u)}{x_e} = \sum_{e\in \Out(u)}{x_e}  &\ & u\in \Vscc \label{eq:cons}\\
& \sum_{e\in \Escc} x_e\cdot \Weight(e) \leq  \vec{\nu} \nonumber\\
& \sum_{e\in \Escc}x_e\geq 1\nonumber
 \end{align}
The quantities $x_e$ are intuitively interpreted as "flows". The first 
constraint specifies that the flow of each edge is non-negative. The second 
constraint is a flow-conservation constraint. The third constraint specifies 
that  the objective is satisfied if we consider the relative contribution of 
the weight of each edge, according to the flow of the edge. The last 
constraint asks that the preceding constraints are satisfied by a non-trivial 
(positive) flow.
Hence, when $\Dim>1$, the decision problem reduces to solving a LP, 
and the time complexity is polynomial~\cite{Khachiyan79}. 

\smallskip\noindent{\em Witness construction.}
The witness path construction from a feasible solution consists of two steps:
(A)~Construction of a multi-cycle from the feasible solution; and 
(B)~Construction of an infinite witness path from the multi-cycle.
We describe the two steps in detail.
Formally, a \emph{multi-cycle} is a finite set of cycles with multiplicity
 $\mathcal{MC}=\{(C_1, m_1), (C_2, m_2),\dots, (C_k, m_k)\}$, such that 
every $C_i$ is a simple cycle and $m_i$ is its multiplicity. 
The construction of a multi-cycle from a feasible solution $\vec{x}$ is as follows.
Let $\mathcal{E}=\{ e : x_e >0 \}$. 
By scaling each edge flow $x_e$ by a common factor $z$, we construct the set 
$\mathcal{X}=\{(e, z\cdot x_e): e \in \mathcal{E}\}$, 
with $\mathcal{X}\subset \Escc\times \Natsplus$. 
Then, we start with $\mathcal{MC}=\emptyset$ and apply iteratively the 
following procedure until $\mathcal{X}=\emptyset$: 
(i)~find a pair $(e_i, m_i)=\arg\min_{(e_j, m_j)\in \mathcal{X}}m_j$,
(ii)~form a cycle $C_i$ that contains $e_i$ and only edges that appear in 
$\mathcal{X}$ (because of Equation~(\ref{eq:cons}), this is always possible), 
(iii)~add the pair $(C_i, m_i)$ in the multi-cycle 
$\mathcal{MC}$, (iv)~subtract $m_i$ from all elements $(e_j,m_j)$ of $\mathcal{X}$ such that 
the edge $e_j$ appears in $C_i$, (v)~remove from $\mathcal{X}$ all $(e_j, 0)$ pairs, 
and repeat.
Since $\Vscc$ is an SCC, there is a path $C_i\rightsquigarrow C_j$ for all $C_i,C_j$ in $\mathcal{MC}$. 
Given the multi-cycle $\mathcal{MC}$, the infinite path that achieves the weight at most $\vec{\nu}$ 
is not periodic, but generated by Procedure~\ref{algo:multi-witness}.

\begin{algorithm}
\small
\SetAlgoNoLine
\SetAlgorithmName{Procedure}{procedure}{List of procedures}
\setstretch{1.05}
\caption{Multi-objective witness}\label{algo:multi-witness}
\KwIn{A graph $G=(V,E)$, and a multi-cycle $\mathcal{MC}=\{(C_1, m_1), (C_2, m_2),\dots, (C_k, m_k)\}$} 
\KwOut{An infinite path $\rho\in \MPI(\Weight,\vec{\nu})$}
\BlankLine
$\ell\leftarrow 1$\\
\While{True}{
Repeat $C_1$ for $\ell\cdot m_1$ times\\
$C_1\rightsquigarrow C_2$\\
Repeat $C_2$ for $\ell\cdot m_2$ times\\
$\dots$\\
Repeat $C_k$ for $\ell\cdot m_k$ times\\
$C_k\rightsquigarrow C_1$\\
$\ell\leftarrow \ell+1$
}
\end{algorithm}
\setcounter{algocf}{0}

\end{compactenum}

\smallskip\noindent{\em Algorithm for ratio objectives.}
We now consider ratio objectives, and present a reduction to mean-payoff objectives.
Consider the weight functions $\Weight_1$, $\Weight_2$ and the 
threshold vector $\vec{\nu}=\frac{\vec{p}}{\vec{q}}$ as the 
component-wise division of vectors $\vec{p},\vec{q}\in \Nats^{\Dim}$. 
We define a new weight function $\Weight: E\rightarrow \Integers^{\Dim}$ such that for all $e\in E$, 
we have $\Weight(e)=\vec{q}\cdot \Weight_1(e)-\vec{p}\cdot \Weight_2(e)$ (where $\cdot$
denotes component-wise multiplication). 
It is easy to verify that $\Ratio(\Weight_1, \Weight_2, \vec{\nu})=\MPI(\Weight, \vec{\mathbf{0}})$, 
and thus we solve the ratio objective by solving the new mean-payoff objective, 
as described above.

\smallskip\noindent{\em Algorithms for conjunctions of objectives.} 
Finally, we consider the conjunction of a safety, a liveness, and a mean-payoff objective
(note that we have already described a reduction of ratio objectives to mean-payoff objectives).
More specifically, given a weight function $\Weight$, a threshold vector $\vec{\nu}\in \Rationals$, and sets $X,Y\subseteq V$, 
we consider the decision problem for the objective $\Phi=\Safe(X)\cap\Live(Y)\cap\MPI(\Weight,\vec{\nu})$.
The procedure is as follows:
\begin{compactenum}
\item Initially compute $G_X$ from $G$ as in the case of a single safety objective. 
\item Then, perform an SCC decomposition on $G_X$. 
\item For every live SCC $\Vscc$ that is reachable from $s$, 
solve for the mean-payoff objective in $\Vscc$. Return yes, if $\MPI(\Weight,\vec{\nu})$ is satisfied in any such $\Vscc$. 
\end{compactenum}
If the answer to the decision problem is yes, then the witness consists of a live SCC $\Vscc$, 
along with a multi-cycle (resp.\ a cycle for $d=1$). 
The witness infinite path is constructed as in Procedure~\ref{algo:multi-witness}, with the only 
difference that at end of each while loop a live node from $Y$ in the SCC $\Vscc$ is additionally visited.
The time required for the conjunction of objectives is dominated by the time required to solve for 
the mean-payoff objective. 
Figure~\ref{fig:obj1} provides a relevant example.

\begin{example}
Consider the graph in Figure~\ref{fig:obj1}. Starting from node $1$, the mean-payoff-objective $\MPI(\Weight, \vec{\mathbf{0}})$ is satisfied by the multi-cycle $\mathcal{MC}=\{(C_1, 1), (C_2, 2)\}$, with $C_1=((1,2), (2,1))$ and $C_2=((3,5), (5,3))$. A solution to the corresponding LP is $x_{(1,2)}=x_{(2,1)}=\frac{1}{3}$ and $x_{(3,5)}=x_{(5,3)}=\frac{2}{3}$, and $x_e=0$ for all other $e\in E$. Procedure~\ref{algo:multi-witness} then generates a witness path for the objective.  The objective is also satisfied in conjuction with $\Safe(\{4\})$ or $\Live(\{4\})$. In the latter case, a witness path additionally traverses the edges $(3,4)$ and $(4,5)$ before transitioning from $C_1$ to $C_2$.
\end{example}

Theorem~\ref{them:graph_objectives} summarizes the results of this section.

\begin{theorem}\label{them:graph_objectives}
Let $G=(V, E)$ be a graph, $s\in V$, $X,Y\subseteq V$, $\Weight:E\rightarrow \Integers^{\Dim}$, $\Weight_1$, $\Weight_2$ $E\rightarrow \Nats^{\Dim}$ weight functions, and $\vec{\nu}\in \Rationals^{\Dim}$. Let $\Phi_1=\Safe(X)\cap\Live(Y)\cap\MPI(\Weight,\vec{\nu})$ and $\Phi_2=\Safe(X)\cap\Live(Y)\cap\Ratio(\Weight_1, \Weight_2,\vec{\nu})$. 
The decision problem of whether $G$ satisfies the objective $\Phi_1$ (resp.\ $\Phi_2$) 
from $s$ requires
\begin{compactenum}
\item $O(n\cdot m)$ time, if $\Dim=1$.
\item Polynomial time, if $\Dim>1$.
\end{compactenum}
If the objective $\Phi_1$ (resp.\ $\Phi_2$) is satisfied in $G$ from $s$, 
then a finite witness (an SCC and a cycle for single dimension, and 
an SCC and a multi-cycle for multiple dimensions) exists and can be constructed in polynomial time.
\end{theorem}

\section{Reduction}\label{sec:reduction}
We present a formal reduction of the computation of 
the competititve ratio of an on-line scheduling algorithm with 
constraints on job sequences to the multi-objective graph problem.
The input consists of the taskset, a deterministic LTS for the 
on-line algorithm, and optional deterministic LTSs for the constraints.

\smallskip\noindent{\em Reduction.}
We first describe the process of computing the competitive ratio $\CR_{\J}(\A)$ where $\J$ is a set of job sequences only subject to safety and liveness constraints. We later show how to handle limit-average constraints.

Given the deterministic and non-deterministic LTS $L_{\A}$ and $L_{\C}$ with reward functions $r_{\A}$ and $r_{\C}$, respectively, and optionally safety and liveness LTS $L_{\S}$ and $L_{\L}$, let $L=L_{\A}\times L_{\C}\times L_{\S}\times L_{\L}$ be their synchronous product. Hence, $L$ is a non-deterministic LTS $(S, s_1,\InAct, \OutAct,\trans)$, and every job sequence $\sigma$ yields a set of runs $R$ in $L$, such that each $\rho\in R$ captures the joint behavior of $\A$ and $\C$ under $\sigma$. Note that for each such $\rho$ the behavior of $\A$ is unchanged, but the behavior of $\C$ generally varies, due to non-determinism. Let $G=(V,E)$ be the multi-graph induced by $L$, that is, $V=S$ and $(M,M', j)\in E$ for all $1\leq j \leq i$ iff there are $i$ transitions $(M, T, M', P)\in \trans$. Let $\Weight_{\A}$ and $\Weight_{\C}$ be the weight functions that assign to each edge of $G$ the reward that the respective algorithm obtains from the corresponding transition in $L$. Let $X$ be the set of states in $G$ whose $L_{\S}$ component is $s_r$, and $Y$ the set of states in $G$ whose $L_{\L}$ component is $s_a$. It follows that for all $\nu\in \Rationals$, we have that $\CR_{\J}(\A)\leq  \nu$ iff the objective $\Phi_{\nu}=\Safe(X)\cap\Live(Y)\cap\Ratio(\Weight_{\A}, \Weight_{\C}, \nu)$ is satisfied in $G$ from the state $s_1$. As the dimension in the ratio objective is one, Case~1 of Theorem~\ref{them:graph_objectives} applies, and we obtain the following:

\begin{lemma}\label{lem:comp_ratio_reduction_one}
Given the product graph $G=(V, E)$ of $n$ nodes and $m$ edges, a rational $\nu\in \Rationals$, and a set of job sequences $\J$ admissible to safety and liveness LTSs, determining whether $\CR_{\J}(\A)\leq \nu$ requires $O(n\cdot m)$ time.
\end{lemma}
Since $0\leq \CR_{\J}(\A) \leq 1$, the problem of determining the competitive ratio reduces to finding $v=\sup \{\nu\in \Rationals:~\Phi_{\nu} \text{ is satisfied in } G\}$. Because this value corresponds to the ratio of the corresponding rewards obtained in a simple cycle in $G$, it follows that $v$ is the maximum of a finite set, and can be determined exactly by a binary search.
Algorithm $\AdaptiveBinarySearch$ (Algorithm \ref{algo:bin_search}) implements an \emph{adaptive} binary search for the competitive ratio in the interval $[0,1]$. The algorithm maintains an interval $[\ell, r]$ such that $\ell\leq \CR_{\J}(\A)\leq r$ at all times, and exploits the nature of the problem for refining the
interval as follows: First, if the current objective $\nu\in [\ell, r]$ (typically, $\nu=(\ell+r)/2$) is satisfied in $G$ i.e., Lemma~\ref{lem:comp_ratio_reduction_one} answers ``yes'' and provides the current minimum cycle $C$ as a witness, the value $r$ is updated to the ratio $\nu'$ of the on-line and off-line rewards in $C$, which is typically less than $\nu$. This allows to reduce the current interval for the next iteration from $[\ell,r]$ to $[\ell, \nu']$, with $\nu'\leq \nu$, rather than $[\ell, \nu]$ (as a simple binary search would do). Second, since $\CR_{\J}(\A)$ corresponds to the ratio of rewards on a simple cycle in $G$, if the current objective $\nu\in [\ell, r]$ is not satisfied in $G$, the algorithm assumes that $\CR_{\J}(\A)=r$ (i.e, the competitive ratio equals the right endpoint of the current interval), and tries $\nu=r$ in the next iteration. Hence, as opposed to a naive binary search, the adaptive version has the advantages of (i)~returning the exact value of $\CR_{\J}(\A)$ (rather than an approximation), and (ii)~being faster.

\smallskip
\begin{algorithm}
\small
\SetAlgoNoLine
\DontPrintSemicolon
\setstretch{1.05}
\caption{$\AdaptiveBinarySearch$}\label{algo:bin_search}
\KwIn{Graph $G=(V,E)$ and weight functions $\Weight_{\A},~\Weight_{\C}$} 
\KwOut{$\min_{C\in G}\frac{\Weight_{\A}(C)}{\Weight_{\C}(C)}$}
\BlankLine
$\ell\leftarrow 0$,
$r\leftarrow 1$,
$\nu\leftarrow \frac{(\ell+r)}{2}$\\
\While{True}{
Solve $G$ for obj. $\Phi_{\nu}$ and find min simple cycle $C$\\
$\nu_1\leftarrow \Weight_{\A}(C)$,
$\nu_2\leftarrow \Weight_{\C}(C)$\\
\eIf{$\nu=\frac{v_1}{v_2}$}{
\Return{$\nu$}
}{
\eIf{$\nu>\frac{v_1}{v_2}$}{
$r\leftarrow\frac{\nu_1}{\nu_2}$,
$\nu\leftarrow\frac{(\ell+r)}{2}$
}
{
$\ell\leftarrow \nu$,
$r\leftarrow\min\left(\frac{\nu_1}{\nu_2}, r\right)$,
$\nu\leftarrow r$
}}
}
\end{algorithm}
 

Finally, we turn our attention to limit-average constraints and the LTS $L_{\W}$. We follow a similar approach as above, but this time including $L_{\W}$ in the synchronous product, i.e., $L=L_{\A}\times L_{\C}\times L_{\S}\times L_{\L}\times L_{\W}$. Let $\Weight_{\A}$ and $\Weight_{\C}$ be weight functions that assign to each edge $e\in E$ in the corresponding multi-graph a vector of $\Dim+1$ weights as follows. In the first dimension, $\Weight_{\A}$ and $\Weight_{\C}$ are defined as before, assigning to each edge of $G$ the corresponding rewards of $\A$ and $\C$. In the remaining $\Dim$ dimensions, $\Weight_{\C}$ is always $1$, whereas $\Weight_{\A}$ equals the value of the weight function $\Weight$ of $L_{\W}$ on the corresponding transition.
Let $\vec{\lambda}$ be the threshold vector of $L_{\W}$. It follows that for all $\nu\in \Rationals$, we have that $\CR_{\J}(\A)\leq  \nu$ iff the objective $\Phi_{\nu}=\Safe(X)\cap\Live(Y)\cap\Ratio(\Weight_{\A}, \Weight_{\C}, (\nu, \vec{\lambda}))$ is satisfied in $G$ from the state $s$ that corresponds to the initial state of each LTS, where $(\nu,\vec{\lambda})$ is a $d+1$-dimension vector, with $\nu$ in the first dimension, followed by the $d$-dimension vector $\vec{\lambda}$. As the dimension in the ratio objective is greater than one, Case~2 of Theorem~\ref{them:graph_objectives} applies, and we obtain the following:

\begin{lemma}\label{lem:comp_ratio_reduction_many}
Given the product graph $G=(V, E)$ of $n$ nodes and $m$ edges, a rational $\nu\in \Rationals$, and a set of job sequences $\J$ admissible to safety, liveness, and limit average LTSs, determining whether $\CR_{\J}(\A)\leq \nu$ requires polynomial time.
\end{lemma}

Again, since $0\leq \CR_{\J}(\A) \leq 1$, the competitive ratio is determined by an adaptive binary search, similar to Algorithm~\ref{algo:bin_search}. 
However, this time $\CR_{\J}(\A)$ is not guaranteed to be realized by a simple cycle (the witness path in $G$ is not necessarily periodic, see Procedure~\ref{algo:multi-witness}), and 
is only approximated within some desired error threshold $\epsilon>0$.

\section{Optimized Reduction}\label{sec:optimization}
In Section~\ref{sec:reduction}, we have established a formal reduction from determining the competitive ratio of an on-line scheduling algorithm in a constrained adversarial environment to solving multiple objectives on graphs. In the current section, we present several optimizations in this reduction that significantly reduce the size of the generated LTSs.

\smallskip\noindent{\em Clairvoyant LTS.} Recall the clairvoyant LTS $L_{\C}$ with reward function $r_{\C}$ from Section~\ref{sec:lts} that non-deterministically models a scheduler. Now we encode the off-line algorithm as a non-deterministic LTS $L_{\C}'=(S_{\C}', s_{\C}',\InAct, \emptyset,\trans_{\C}')$ with reward function $r_{\C}'$ that lacks the property of being a scheduler, as information about released and scheduled jobs is lost. However, it preserves the property that, given a job sequence $\sigma$, there exists a run $\rho_{\C}^{\sigma}$ in $L_{\C}$  iff there exists a run $\widehat{\rho}_{\C}^{\sigma}$ in $L_{\C}'$ with  $V(\rho_{\A}^{\sigma},k)=V(\widehat{\rho}_{\A}^{\sigma},k)$ for all $k\in \Natsplus$. That is, there is a bisimulation between $L_{\C}$ and $L_{\C}'$ that preserves rewards.

Intuitively, the clairvoyant algorithm need not partially schedule a task, i.e., it will either discard it immediately, or schedule it to completion. Hence, in every release of a set of tasks $T$, $L_{\C}'$ non-deterministically chooses a subset $T'\subseteq T$ to be scheduled, as well as allocates the future slots for their execution. Once these slots are allocated, $L_{\C}'$ is not allowed to preempt those in favor of a subsequent job. 

The state space $S_{\C}'$ of $L_{\C}'$ consists of binary strings of length $D_{\max}$. For a binary string $B\in S_{\C}'$, we have $B[i]=1$ iff the $i$-th slot in the future is allocated to some released job, and $s_{\C}'=\vec{\mathbf{0}}$. Informally, the transition relation $\trans_{\C}'$ is such that, given a current subset $T\subseteq \Sigma$ of released jobs, there exists a transition $\delta$ from $B$ to $B'$ only if $B'$ can be obtained from $B$ by non-deterministically choosing a subset $T'\subseteq T$, and for each task $\tau_i\in T'$ allocating non-deterministically $C_i$ free slots in $B$.
Finally, set $r_{\C}'=\sum_{\tau_i\in T'}V_i$. 

By definition, $|S_{\C}'|\leq 2^{D_{\max}}$. In laxity-restricted tasksets, we can obtain an even  tighter bound. Let $L_{\max}=\max_{\tau_i\in \Tau}{(D_i-C_i)}$ be the maximum laxity in $\Tau$, and $I:S_{\C}'\rightarrow \{\bot,1,\dots,D_{\max}-1\}^{L_{\max}+1}$ a function such that $I(B)=(i_1,\dots, i_{L_{\max}+1})$ are the indexes of the first $L_{\max}+1$ zeros in $B$. That is, $i_j=k$ iff $B[k]$ is the $j$-th zero location in $B$, and $i_j=\bot$ if there are less than $j$ free slots in $B$.

\begin{claim}\label{claim:laxity}
The function $I$ is bijective.
\end{claim}

\begin{proof} 
Fix a tuple $(i_1,\dots,i_{L_{\max}+1})$, and let $B\in S_{\C}'$ be any state such that $I(B)=(i_1,\dots,i_{L_{\max}+1})$. We consider two cases.
\begin{enumerate}
\item If $i_{L_{\max}+1}=\bot$, there are less than $L_{\max}+1$ empty slots in $B$, all uniquely determined by $(i_1,\dots,i_{k})$, for some $k\leq L_{\max}$.
\item If $i_{L_{\max}+1}\neq \bot$, then all $i_j\neq \bot$, and thus any job to the right of $i_{L_{\max}+1}$ would have been stalled for more than $L_{\max}$ positions. Hence, all slots to the right of $i_{L_{\max}+1}$ are free in $B$, and $B$ is also unique.
\end{enumerate}
Hence, $I(B)$ always uniquely determines $B$, as desired.
\end{proof}

For $x,k\in \Natsplus$, denote with $\Perm(x,k)=x\cdot(x-1) \dots (x-k+1)$ the number of $k$-permutations on a set of size $x$. 
\begin{lemma}\label{lem:offline_statespace}
Let $\Tau$ be a taskset with maximum deadline $D_{\max}$, and $L_{\max}=\max_{\tau_i\in \Tau}{(D_i-C_i)}$ be the maximum laxity. Then, $|S_{\C}'|\leq \min(2^{D_{\max}}, \Perm(D_{\max},L_{\max}+1))$.
\end{lemma}

Hence, for zero and small laxity environments~\cite{BKMM92}, as e.g. arising in wormhole switching in NoCs~\cite{LJ07}, $S_{\C}'$ has polynomial size in $D_{\max}$.

\smallskip\noindent{\em Clairvoyant LTS generation.}
We now turn our attention on efficiently generating the clairvoyant LTS $L_{\C}'$ as described in the previous paragraph. There is non-determinism in two steps: both in choosing the subset $T'\subseteq T$ of the currently released tasks for execution, and in allocating slots for executing all tasks in $T'$. Given a current state $B$ and $T$, this non-determinism leads to several identical transitions $\delta$ to a state $B'$. We have developed a recursive algorithm called $\ClairvoyantSuccessor$ (Algorithm~\ref{algo:clairvoyant_successor}) that generates each such transition $\delta$ exactly once.

The intuition behind $\ClairvoyantSuccessor$ is as follows. It has been shown that the earliest deadline first (EDF) policy is optimal in scheduling job sequences where every released task can be completed~\cite{Dertouzos74}. By construction, given a job sequence $\sigma_1$,  $L_{\C}'$ non-deterministically chooses a job sequence $\sigma_2$, such that for all $\ell$, we have $\sigma_2^{\ell}\subseteq \sigma_1^{\ell}$, and all jobs in $\sigma_2$ are scheduled to completion by $L_{\C}'$. Therefore, it suffices to consider a transition relation $\trans_{\C}'$ that allows at least all possible choices that admit a feasible EDF schedule on every possible $\sigma_2$, for any generated job sequence $\sigma_1$.

In more detail, $\ClairvoyantSuccessor$ is called with a current state $B$, a subset of released tasks $T$ and an index $k$, and returns the set $\mathcal{B}$ of all possible successors of $B$ that schedule a subset $T'\subseteq T$, and every job of $T'$ is executed later than $k$ slots in the future. This is done by extracting from $T$ the task $\tau$ with the earliest deadline, and proceeding as follows: The set $\mathcal{B}$ is obtained by constructing a state $B'$ that considers all the possible ways to schedule $\tau$ to the right of $k$ (including the possibility of not scheduling $\tau$ at all), and recursively finding all the ways to schedule $T\setminus\{\tau\}$ in $B'$, to the right of the rightmost slot allocated for task $\tau$.

Finally, we exploit the following two observations to further reduce the state space of $L_{\C}'$. First, we note that as long as there is some load in the state of $L_{\C}'$ (i.e., at least one bit of $B$ is one), the clairvoyant algorithm gains no benefit by not executing any job in the current slot. Hence, besides the zero state $\vec{\mathbf{0}}$, every state $B$ must have $B[1]=1$. In most cases, this restriction reduces the state space by at least $50\%$. Second, it follows from our claims on the off-line EDF policy of the clairvoyant scheduler that for every two scheduled jobs $J$ and $J'$, it will never have to preempt $J$ for $J'$ and vice versa. A consequence of this is that, for every state $B$ and every continuous segment of zeros in $B$ that is surrounded by ones (called a \emph{gap}), the gap must be able to be completely filled with some jobs that start and end inside the gap. This reduces to solving a knapsack problem~\cite{Karp72} where the size of the knapsack is the length of the gap, and the set of items is the whole taskset $\Tau$ (with multiplicities). We note that the problem has to be solved on identical inputs a large number of times, and techniques such as \emph{memoization} are employed to avoid multiple evaluations of the same input. 

These two improvements were found to reduce the state space by a factor up to $90\%$ in all examined cases (see Section~\ref{sec:results} and Table~\ref{table:runtimes}), and despite the non-determinism, in all reported cases the generation of $L_{\C}$ was done in less than a second.

\smallskip
\begin{algorithm}
\small
\SetAlgoNoLine
\DontPrintSemicolon
\setstretch{1.05}
\caption{$\ClairvoyantSuccessor$}\label{algo:clairvoyant_successor}
\KwIn{A set $T\subseteq \Tau$, state $B$, index $1\leq k\leq D_{\max}$} 
\KwOut{A set $\mathcal{B}$ of successor states of $B$ }
\BlankLine
\lIf{$T=\emptyset$}{
\Return $\{B\};$
}
$\tau\leftarrow \arg\min_{\tau_i\in T}D_i$, $C\leftarrow \text{ execution time of }\tau$\\
$T'\leftarrow T\setminus \{\tau\}$\\
\tcp{Case 1: $\tau$ is not scheduled}
$\mathcal{B}\leftarrow\ClairvoyantSuccessor(T', B, k)$\\
\tcp{Case 2: $\tau$ is scheduled}
$\mathcal{F}\leftarrow \text{set of free slots in } B \text{ greater than } k$\\
\ForEach{$F\subseteq \mathcal{F}$ with $|F|=C$}{
$B'\leftarrow \text{Allocate } F \text { in } B$\\
$k'\leftarrow \text{rightmost slot in } F$\\
$\mathcal{B}'\leftarrow \ClairvoyantSuccessor(T', B', k')$\\
\tcp{Keep only non-redundant states}
\ForEach{$B''\in \mathcal{B}'$}{
\If{$B''[1]=1$ and $\Knapsack(B'',\Tau)$}{
$\mathcal{B}\leftarrow \mathcal{B} \cup \{B''\}$
}
}
}
\Return $\mathcal{B}$
\end{algorithm}

\smallskip\noindent{\em On-line state space reduction.} Typically, most on-line scheduling algorithms do ``lazy dropping'' of the jobs, where a job is dropped only when its deadline passes. To keep the
state-space of the LTS small, it is crucial to only store those jobs that have the possibility of being scheduled, at least partially, under some sequence of future task releases. We do so by first creating the LTS naively, and then iterating through its states. For each state $s$ and job $J_{i,j}$ in $s$ with relative deadline $D_{i}$, we perform a \emph{depth-limited search} originating in $s$ for $D_{i}$ steps, looking for a state $s'$ reached by a transition that schedules $J_{i,j}$. If no such state is found, we merge state $s$ to $s''$, where $s''$ is identical to $s$ without job $J_{i,j}$.

\section{Experimental Results}\label{sec:results}
We have implemented our approach for automated competitive ratio analysis, 
and applied it to a range of case studies: 
four well-known scheduling policies, namely, EDF (Earliest Deadline First), 
SRT (Shortest Remaining Time), SP (Static Priorities), and FIFO (First-in First-out), as well as 
some more elaborate algorithms that provide 
non-trivial performance guarantees, in particular, DSTAR~\cite{Baruah91},  DOVER~\cite{KS95} and  TD1~\cite{BKMM92},
 are analyzed under a variety of tasksets.
Our implementation is done in Python and C, and uses the lp\_solve~\cite{lpsolve} 
package for linear programming solutions.
All experiments are run on a standard 2010 computer with a 3.2GHz CPU and 4GB of RAM running Linux.

\begin{figure*}
\centering
\includegraphics[scale=1]{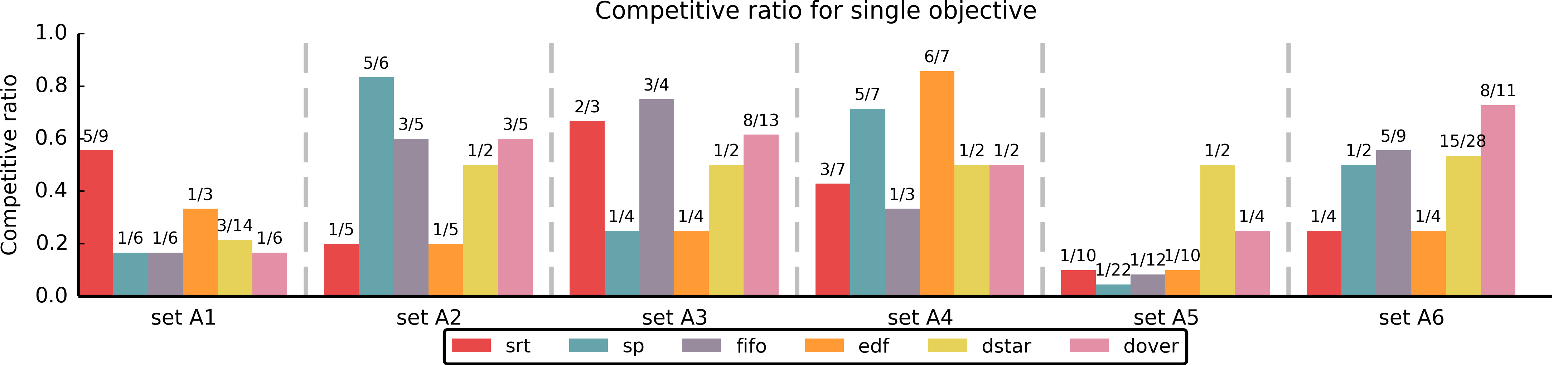}
\caption{The competitive ratio of the examined algorithms in various tasksets under no constraints; the tasksets A1-A6 are available in the Appendix Section~\ref{sec:tasksets}. 
Every examined algorithm is optimal in some taskset, among all others.
}\label{fig:results_single}
\end{figure*}

\smallskip\noindent{\em Varying tasksets without constraints.}
The algorithm DOVER was proved in \cite{KS95} to have optimal competitive 
factor, i.e., optimal competitive ratio under the worst-case taskset. 
However, our experiments reveal that this performance guarantee is 
not universal, in the sense that DOVER is outperformed by other schedulers 
for specific tasksets. 
This observation applies to all on-line algorithms examined: As shown in Figure~\ref{fig:results_single}, 
even without constraints on the adversary, for every scheduling algorithm, there are tasksets 
in which it achieves the highest competitive ratio among all others. Note that this
high variability of the optimal on-line algorithm across tasksets 
makes our automated analysis framework an interesting tool for the 
application designer.

\begin{figure}
\centering
\includegraphics[scale=1]{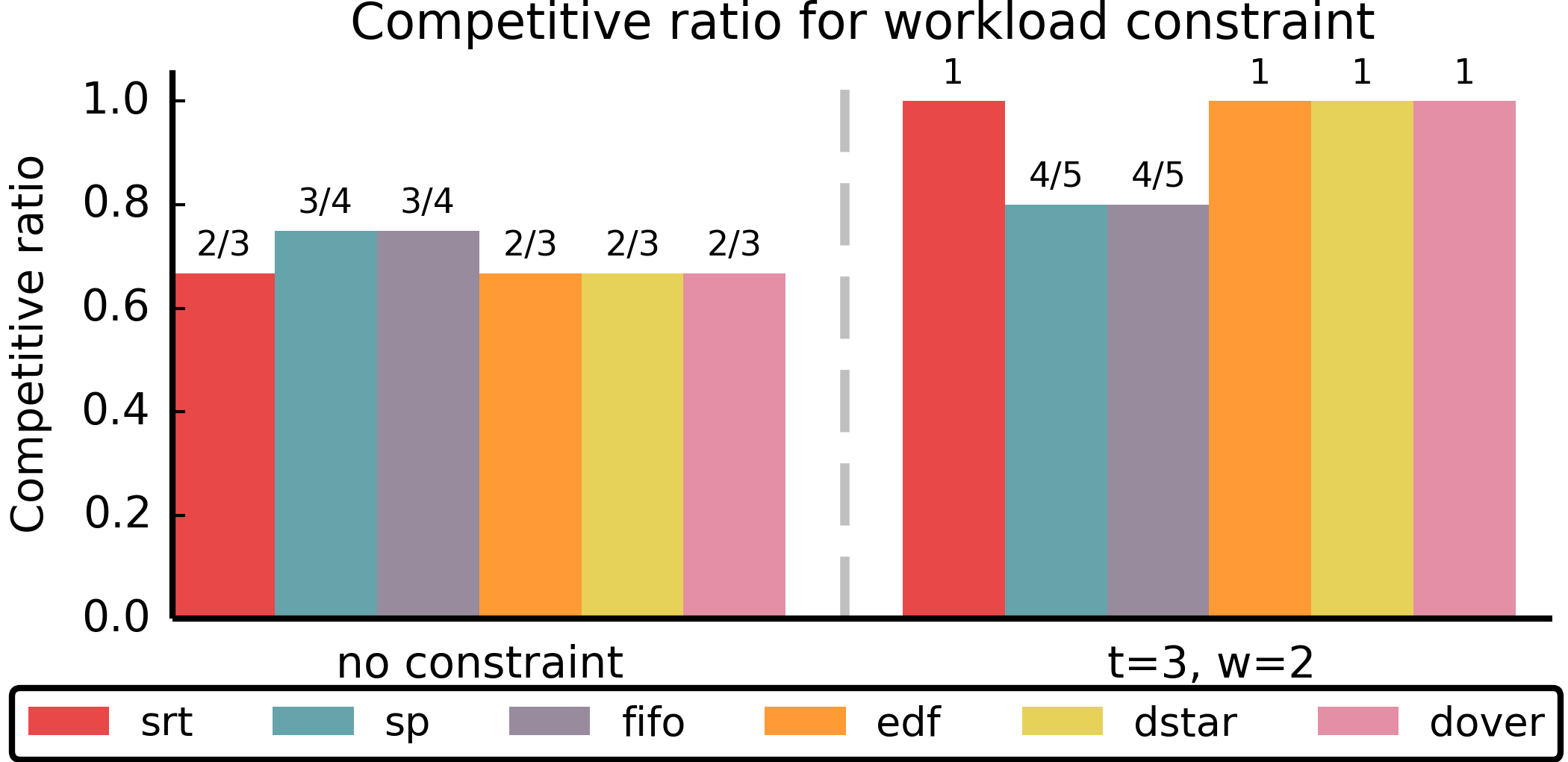}
\caption{Restricting the absolute workload generated by the adversary typically increases the competitive ratio, and can vary the optimal scheduler. On the left, the performance of each scheduler is evaluated without restrictions: FIFO, SP behave best. When restricting the adversary to at most $2$ units of workload in the last $3$ rounds, FIFO and SP become suboptimal, and are outperformed by other schedulers.
The taskset is available in the Appendix Section~\ref{sec:tasksets}.} \label{fig:abs_workload}
\end{figure}

\smallskip\noindent{\em Fixed taskset with varying constraints.}
We also consider fixed tasksets under various constraints 
(such as sporadicity or workload restrictions) for admissible job sequences.
%
Figure~\ref{fig:abs_workload} shows our experimental results for workload safety constraints, 
which again reveal that, depending on workload constraints, we can have different optimal schedulers. 
Finally, we consider limit-average constraints and observe that varying these constraints can also vary the optimal scheduler for a fixed taskset:
As Table~\ref{table:workload} shows, the optimal scheduler can vary highly and non-monotonically with stronger limit-average workload restrictions. 

\begin{table}[!h]
\small
\centering
\begin{tabular}{|c|| c | c | c | c | c | c | c | c | c | }
\hline
 & $1.5$ & $1$ & $0.8$ & $0.6$ & $0.4$ & $0.3$ & $0.1$ & $0.78$ & $0.05$\\
 \hline
 \hline
 fifo & \checkmark & \checkmark & \checkmark & \checkmark & \checkmark &&&& \checkmark \\
 \hline
 sp & \checkmark &&&&&& \checkmark &&\checkmark \\
 \hline
 srt & \checkmark &&&&\checkmark & \checkmark & \checkmark  & \checkmark & \checkmark\\
 \hline
\end{tabular}
\caption{Columns show the mean workload restriction. The check-marks indicate that the corresponding scheduler is optimal for that mean workload restriction, among the six schedulers we examined. We see that the optimal scheduler can vary as the restrictions are tighter, and in a non-monotonic way. EDF, DSTAR and DOVER were not optimal in any case and hence not mentioned.
The taskset is available in the Appendix Section~\ref{sec:tasksets}.}
\label{table:workload}
\end{table}

\smallskip\noindent{\em Competitive ratio of TD1.}
We also consider the performance of the online scheduler TD1
in zero laxity tasksets with uniform value-density (i.e., for each task $\tau_i$,
we have $C_i=D_i=V_i$). Following~\cite{BKMM92},
we construct a series of tasksets parameterized by some positive real $\eta<4$,
which guarantee that the competitive ratio
of every online scheduler is upper bounded by $\frac{1}{\eta}$.
Given $\eta$, each taskset consists of tasks $\tau_i$ such that $C_i$ is given
by the following recurrence, as long as $C_{i+1}>C_i$.

\[
(i)~C_0=1 \qquad (ii)~C_{i+1} = \eta\cdot C_i - \sum_{j=0}^i C_j
\]

In~\cite{BKMM92}, TD1 was shown to have competitive factor $\frac{1}{4}$, and hence
a competitive ratio that approaches $\frac{1}{4}$ from above, as $\eta\rightarrow 4$ in the above series of tasksets.
Table~\ref{table:td1} shows the competitive ratio
of TD1 in this series of tasksets.
Each taskset is represented as a set $\{C_i\}$, where each $C_i$ is given
by the above recurrence, rounded up to the first integer.
We indeed see that the competitive ratio drops
until it stabilizes to $\frac{1}{4}$. 

Finally, note that the zero-laxity restriction allows us to process tasksets 
where $D_{\max}$ is much higher than what we report in Table~\ref{table:runtimes}.
The results of Table~\ref{table:td1} are produced in less than a minute in total.

\begin{table}[!h]
\small
\centering
\begin{tabular}{| c | c | c || c |}
\hline
Name & $\eta$ & Taskset & Comp. Ratio\\
\hline
\hline
set C1 & $2$ & $\{1,1\}$ & $\mathbf{1}$\\
\hline
set C2 & $3$ & $\{1,2,3\}$ & $\mathbf{1/2}$\\
\hline
set C3 & $3.1$ & $\{1,3,7,13,19\}$ & $\mathbf{7/25}$\\
\hline
set C4 & $3.2$ & $\{1,3,7,13,20,23\}$ & $\mathbf{1/4}$\\
\hline
set C5 & $3.3$ & $\{1,3,7,14,24,33\}$ & $\mathbf{1/4}$\\
\hline
set C6 & $3.4$ & $\{1,3,7,14,24,34\}$ & $\mathbf{1/4}$\\
\hline
\end{tabular}
\caption{Competitive ratio of TD1}
\label{table:td1}
\end{table}

\smallskip\noindent{\em Running times.}
Table~\ref{table:runtimes} summarizes some key parameters of our various tasksets, and gives
some statistical data on the observed running times in our respective experiments.
Even though the considered tasksets are small, the very short running
times of our prototype implementation 
reveal the principal feasibility of our approach. 
We believe that further application-specific optimizations, augmented by 
abstraction and symmetry reduction techniques, will allow to scale to larger 
applications.

%

\begin{table}[!h]
\small
\centering
\begin{tabular}{| c || c | c | c | c | c | c |}
\hline
Name & N & $D_{\max}$ & \multicolumn{2}{c|}{Size (nodes)} & \multicolumn{2}{c|}{Time (s)} \\
\hline
&&& Clairv. & Product & Mean & Max\\
\hline
\hline
set B01 & $2$ & $7$ & $19$ & $823$ & $0.04$ & $0.05$\\
\hline
set B02 & $2$ & $8$ & $26$ & $1997$ & $0.39$ & $0.58$\\
\hline
set B03 & $2$ & $9$ & $34$ & $4918$ & $10.02$ & $15.21$\\
\hline
set B04 & $3$ & $7$ & $19$ & $1064$ & $0.14$ & $0.40$\\
\hline
set B05 & $3$ & $8$ & $26$ & $1653$ & $0.66$ & $2.05$\\
\hline
set B06 & $3$ & $9$ &$34$ & $7705$ & $51.04$ & $136.62$\\
\hline
set B07 & $4$ & $7$ & $19$ & $1711$ &  $2.13$ & $6.34$\\
\hline
set B08 & $4$ & $8$ & $26$ & $3707$ & $13.88$ & $34.12$\\
\hline
set B09 & $4$ & $9$ & $44$ & $10040$ & $131.83$ & $311.94$\\
\hline
set B10 & $5$ & $7$ & $19$ & $2195$ & $5.73$ & $16.42$\\
\hline
set B11 & $5$ & $8$ & $32$ & $9105$ & $142.55$ & $364.92$\\ 
\hline
set B12 & $5$ & $9$ & $44$ & $16817$ & $558.04$ & $1342.59$\\
\hline
\end{tabular}
\caption{Scalability of our approach for tasksets of various sizes $N$
and $D_{\max}$. For each taskset, the size of the state space of the clairvoyant scheduler
is shown, along with the mean size of the product LTS, and the mean and maximum
time to solve one instance of the corresponding ratio objective.}
\label{table:runtimes}
\end{table}

\section{Conclusions}\label{sec:conclusions}
We presented a flexible framework for automatically 
analyzing the competitive ratio 
of on-line scheduling algorithms for an input firm-deadline taskset,
which also supports various forms of constraints for admissible job sequences. 
Our experimental results demonstrate that it allows to solve small-sized 
problem instances efficiently. Moreover, they highlight 
the importance of our fully automated 
approach, as there is neither a ``universally'' 
optimal algorithm for all tasksets (even in the absence of additional 
constraints) nor an optimal algorithm for different constraints in the 
same taskset. 
Thanks to the flexibility of our approach, it can be extended in
various ways (multiple processors, algorithm-specific constraints
like energy restrictions, more general deadlines, etc.). 
Part of our future research will be devoted to incorporating such features. 

\bibliographystyle{unsrt}
\bibliography{lit}

\clearpage
\appendix
\renewcommand{\thesubsection}{\Alph{subsection}}

\subsection{Tasksets used in the reported experiments}\label{sec:tasksets}
Table~\ref{table:set1} lists the tasksets A1-A6 used for Figure~\ref{fig:results_single},
Table~\ref{table:set2} gives the tasksets used in the experiments reported 
in Figure~\ref{fig:abs_workload}
and Table~\ref{table:workload}. In all cases, tasks are ordered by their static priorities,
which determine the SP scheduler, as well as the way ties are broken by other schedulers.
In Table~\ref{table:set1}, along with each tasket, its \emph{importance ratio} $k$ is shown,
defined as
\[
k=\max_{\tau_i,\tau_j\in \Tau} \frac{V_i/C_i}{V_j/C_j}
\]

\begin{table}[!h]
\small
\centering
\begin{tabular}{|c| c | c | c || c ||c| c | c | c |}
\hline
Name & $C_i$ & $D_i$ & $V_i$ & & Name & $C_i$ & $D_i$ & $V_i$ \\
\hline
\hline
set A1 & $1$ & $2$ & $3$ && set A4 & $1$ & $2$ & $3$\\
$k=6$ & $4$ & $6$ & $2$ && $k=3$ & $2$ & $3$ & $2$\\
& $1$ & $3$ & $3$ &&& $1$ & $6$ & $1$\\
\cline{6-9}
& $3$ & $4$ & $3$ && set A5 & $2$ & $2$ & $1$\\
\cline{1-4}
set A2 & $2$ & $3$ & $5$ && $k=4$ & $6$ & $6$ & $10$\\
$k=5$ & $2$ &  $2$ & $1$ &&& $1$ & $1$ & $2$\\
\cline{1-4}
\cline{6-9}
set A3 & $2$ & $2$ & $1$ && set A6 & $1$ & $5$ & $5$\\
$k=4$ & $1$ & $5$ & $2$ && $k=5$ & $2$ & $2$ & $4$\\
& $1$ & $5$ & $2$ &&& $1$ & $1$ & $1$\\
\hline
\end{tabular}
\caption{Tasksets of Figure~\ref{fig:results_single}}
\label{table:set1}
\end{table}

\begin{table}[!h]
\small
\centering
\begin{tabular}{|c | c | c |}
\hline
$C_i$ & $D_i$ & $V_i$\\
\hline
\hline
$1$ & $1$ & $3$\\
$1$ & $2$ & $3$\\
$1$ & $1$ & $1$\\
\hline
\end{tabular}
\hspace*{2cm}
\begin{tabular}{|c | c | c |}
\hline
$C_i$ & $D_i$ & $V_i$\\
\hline
\hline
$2$ & $7$ & $3$\\
$5$ & $5$ & $2$\\
$5$ & $6$ & $1$\\
\hline
\end{tabular}
\caption{Taskset of Figure~\ref{fig:abs_workload} (left) and Table~\ref{table:workload} (right).}
\label{table:set2}
\end{table}


\subsection{Examples of on-line schedulers as LTSs}\label{sec:ex-sch}
\begin{figure}[!h]
\centering
\subfloat[]{
\includegraphics[scale=0.17]{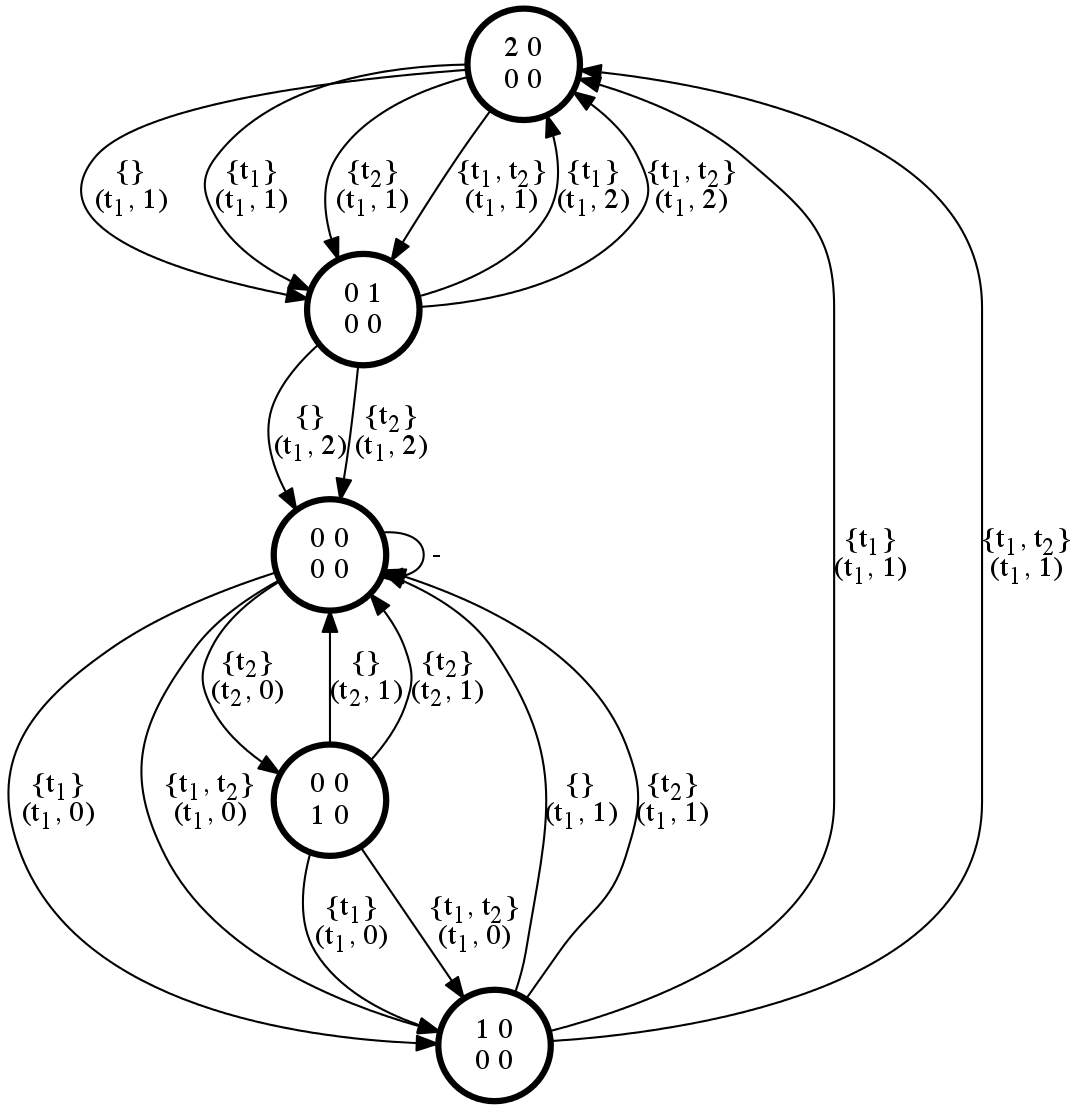}\label{fig:lts_sp_set1}
}\\
\subfloat[]{
\includegraphics[scale=0.17]{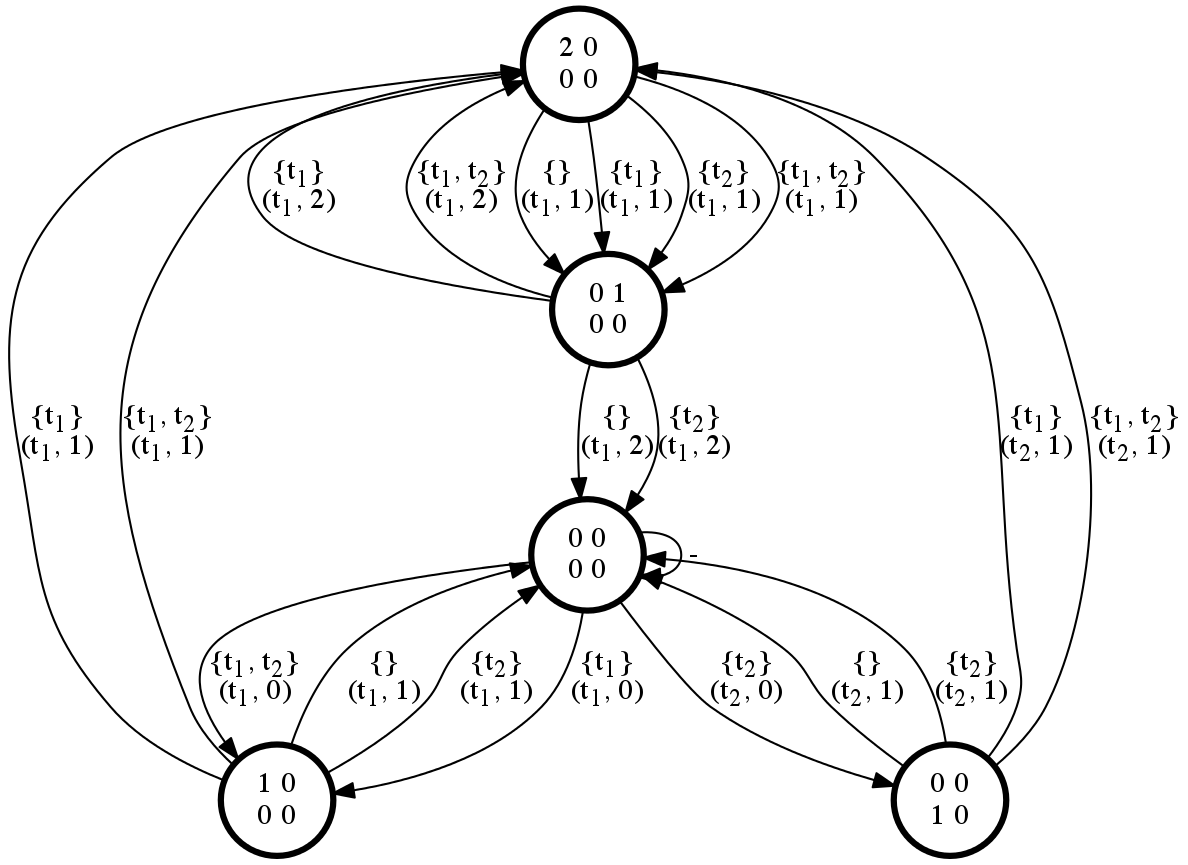}\label{fig:lts_fifo_set1}
}\\
\subfloat[]{
\includegraphics[scale=0.17]{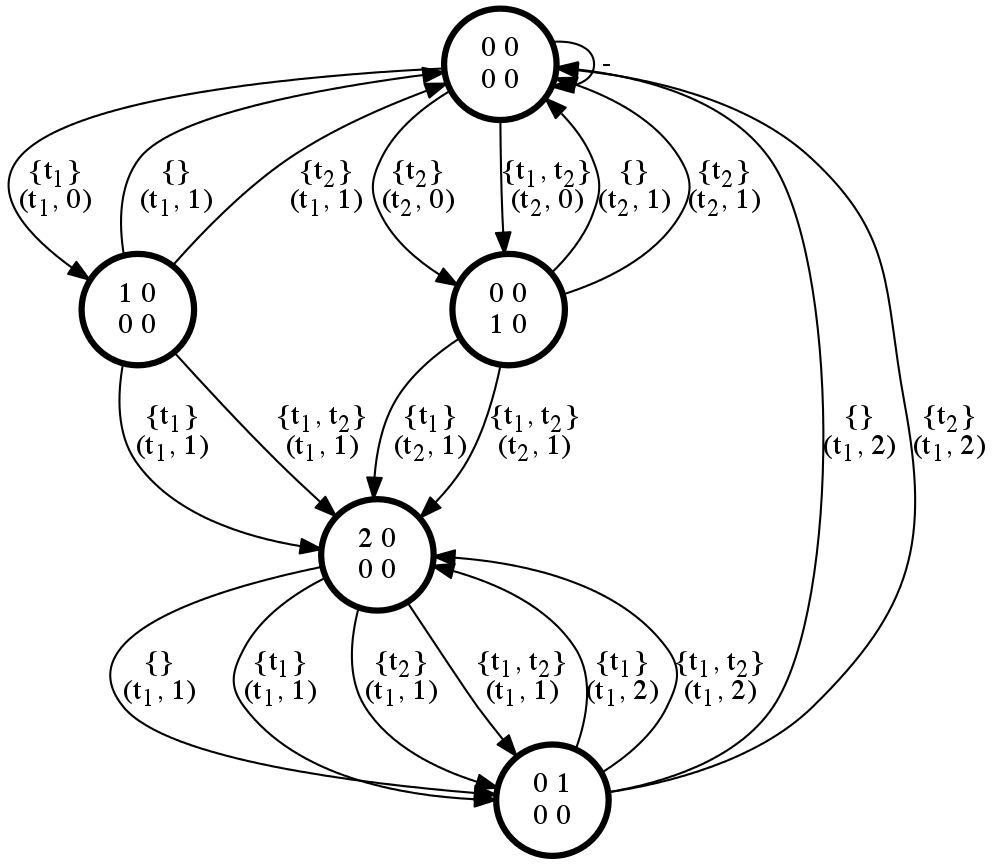}\label{fig:lts_srt_set1}
}
\caption{The SP~\protect\subref{fig:lts_sp_set1},
FIFO~\protect\subref{fig:lts_fifo_set1}  and SRT~\protect\subref{fig:lts_srt_set1}, 
on-line scheduling algorithms for the taskset $\Tau=\{\tau_1, \tau_2\}$ with $D_1=3$, $D_2=2$ and 
$C_1=C_2=1$, represented as LTSs. 
}\label{appfig:lts}
\end{figure}

We now present more examples of on-line schedulers represented as deterministic 
LTSs. 
Consider the taskset $\Tau=\{\tau_1, \tau_2\}$ with $D_1=3$, $D_2=2$ and 
$C_1=C_2=1$ already used for the EDF example in Figure~\ref{ex:edf}.
Figure~\ref{appfig:lts} shows the EDF, SP, FIFO, and SRT policies represented 
as deterministic LTSs.

\end{document}